\documentclass[usenatbib]{mnras}

\usepackage{newtxtext,newtxmath}

\usepackage[T1]{fontenc}

\DeclareRobustCommand{\VAN}[3]{#2}
\let\VANthebibliography\thebibliography
\def\thebibliography{\DeclareRobustCommand{\VAN}[3]{##3}\VANthebibliography}

\usepackage{graphicx}	
\usepackage{amsmath}	
\usepackage{siunitx}
\DeclareSIUnit\angstrom{\text {Å}}
\usepackage{float}
\usepackage{hyperref}
\usepackage{caption}
\usepackage{tabularx}
\usepackage{pdflscape}
\usepackage{afterpage}
\usepackage{lipsum}

\title[The oldest stars from the early universe]
{The oldest stars with low neutron-capture element abundances and origins in ancient dwarf galaxies}

\author[H.D. Andales et al.]{
Hillary Diane Andales,$^{1}$\thanks{E-mail: handales@mit.edu}
Ananda Santos Figueiredo,$^{1}$
Casey Gordon Fienberg,$^{1}$ 
\newauthor~Mohammad K.\ Mardini,$^{2,3,4,1}$
Anna Frebel$^{1,4}$
\\
$^{1}$Department of Physics and Kavli Institute for Astrophysics and Space Research, Massachusetts Institute of Technology, Cambridge, MA 02139, USA\\
$^{2}$Department of Physics, Zarqa University, Zarqa 13110, Jordan\\
$^{3}$Jordanian Astronomical Virtual Observatory, Zarqa University, Zarqa 13110, Jordan\\
$^{4}$Joint Institute for Nuclear Astrophysics – Center for the Evolution of the Elements (JINA-CEE), East Lansing, MI 48824, USA\\
}

\date{Accepted XXX. Received YYY; in original form ZZZ}

\pubyear{2024}

\begin{document}
\label{firstpage}
\pagerange{\pageref{firstpage}--\pageref{lastpage}}
\maketitle

\begin{abstract} 
We present a detailed chemical abundance and kinematic analysis of six extremely metal-poor ($-4.2 \leq$ [Fe/H] $\leq-$2.9) halo stars with very low neutron-capture abundances ([Sr/H] and [Ba/H]) based on high-resolution Magellan/MIKE spectra. Three of our stars have [Sr/Ba] and [Sr/H] ratios that resemble those of metal-poor stars in ultra-faint dwarf galaxies (UFDs). Since early UFDs may be the building blocks of the Milky Way, extremely metal-poor halo stars with low, UFD-like Sr and Ba abundances may thus be ancient stars from the earliest small galactic systems that were accreted by the proto-Milky Way. We label these objects as Small Accreted Stellar System (SASS) stars, and we find an additional 61 similar ones in the literature. 
A kinematic analysis of our sample and literature stars reveals them to be fast-moving halo objects, all with retrograde motion, indicating an accretion origin.
Because SASS stars are much brighter than typical UFD stars, identifying them offers promising ways towards detailed studies of early star formation environments. From the chemical abundances of SASS stars, it appears that the earliest accreted systems were likely enriched by a few supernovae whose light element yields varied from system to system. Neutron-capture elements were sparsely produced and/or diluted, with $r$-process nucleosynthesis playing a role. These insights offer a glimpse into the early formation of the Galaxy. Using neutron-capture elements as a distinguishing criterion for early formation, we have access to a unique metal-poor population that consists of the oldest stars in the universe. 
\end{abstract}





\section{Introduction}

The Milky Way's ultra-faint dwarf satellite galaxies (UFDs) continue to offer a vital window into the early universe. Besides providing insight on a broad range of astrophysical questions such as the formation of galaxies and structure \citep[e.g.,][]{Frebel12, Wheeler2015_FIRE, Kallivayalil2018_LMCUFDs, Applebaum2021_JLSim} and the small-scale properties of dark matter \citep[e.g.,][]{SimonGeha2007_UFDKinematics, Calabrese2016_ULDM, Brandt2016_UFDMachos, Safarzadeh2020_ULDM, Orkney2022_EriIIDM}, UFDs are uniquely important laboratories for studying the early universe. In particular, they provide a glimpse into early star formation environments, early chemical enrichment events, and the early assembly of the Milky Way \citep[e.g,][]{kirby08, Simon2010_LeoIV, Roederer2018_rpUFD, Lee2019_HaloCEMP, Brauer2022_UFDKinematics}. 

Formed 12 to 13 billion years ago at around $z \sim 10$ \citep{Brown2012_HerLeoUMa, Brown14, simon2023}, UFDs are ancient and thus extremely metal-poor \citep{Frebel10a, Frebel10b, ji16c, Chiti18, Chiti2021,chiti23}. In fact, the majority of UFD stars have metallicities between $\mbox{[Fe/H]}=-2$ and $\mbox{[Fe/H]}=-3$, with some stars reaching levels as low as $\mbox{[Fe/H]}=-4$ \citep{Simon2019}. Most of their stars formed before reionization ($z \sim 6$) after which their star formation was quenched \citep{Weisz2014_UFDSFR, Brown14}. Since the epoch of reionization, UFDs have been gas-poor, quiescent, and largely chemically unaltered \citep{Koch09}. As a result, their long-lived metal-poor stars now carry clean signatures of the earliest chemical enrichment events in the universe, as well as hints about early star formation environments. 

However, further progress in exploring the early universe using UFDs is severely hampered by the simple fact that UFDs are too distant. Because these tiny galaxies lie dozens or hundreds of kiloparsecs away, studying their chemical makeup with high-resolution spectroscopy is notoriously challenging. In principle, faint stars with visual magnitudes of $V \sim 19$ are still observable but such observations would be near the technical limits of current telescopes. Thus only a handful of UFD stars — the brightest cool red giants — are available for high-resolution spectroscopy. Much fainter stars are much more difficult to observe, even with medium-resolution spectroscopy. As a result, the current chemical abundance datasets on UFDs are still too sparse and inadequate; only a few dozen UFD stars have detailed chemical abundances \citep{Suda2017_SAGA, Abohalima2018_JINAbase}. To get around this, one common solution is to combine stars from various UFDs and treat that combination as one "population" \citep[e.g.,][]{Vargas13}. Yet, even with this workaround, most of the chemical abundance data still have large uncertainties. 

Unraveling the histories of UFDs requires more detailed probing, much more than the current chemical abundance datasets can allow. We thus suggest an alternative path forward: searching in the Galactic halo for erstwhile member stars of small satellites that were accreted by the Milky Way billions of years ago. After all, the Milky Way assembled hierarchically, and presumably many small "building block" systems — including possible analogs of the surviving UFDs — contributed to its early formation \citep[e.g.,][]{wetzel16, Griffen16}. Stars from these small accreted systems could therefore be suitable analogs of the stars in surviving UFDs. Using these analog stars is obviously advantageous for detailed chemical study since halo stars are more numerous and more accessible for high-resolution spectroscopy.

To isolate these analogs from the broader collection of halo stars, we look for stars with similar chemical properties as those in extant UFDs. First, they must be metal-poor since UFD stars are typically Very Metal-Poor (VMP; $\mbox{[Fe/H]}<-2$) or Extremely Metal-Poor (EMP; $\mbox{[Fe/H]}<-3$). Second, unlike VMP and EMP stars in the halo, we would expect most UFD stars not to show significant carbon enhancements (where carbon enhancement is defined as [C/Fe] $>$ 0.7). This discrepancy is well-established and works in our favor here \citep{Frebel10b}. Third, they must have extraordinarily low abundances of neutron-capture elements (e.g., Sr and Ba). Finally, they must have retrograde motion consistent with an accretion origin. Previous works have used some combination of these criteria to associate halo stars with their possible origins in now-destroyed UFD-like satellites, effectively finding some of our possible analog stars \citep{Hansen2016_CEMPno, Lee2017_CHalo, Yoon2018_AEGISCFe, Yuan20}. Intriguingly, \cite{Sestito24} even found a VMP star in the inner Galaxy with putative origins in an accreted UFD.

Low neutron-capture abundance is the hallmark characteristic of UFD stars that sets them apart from the general halo trend \citep{Ji19}. Many chemical abundance studies have consistently found low $\mbox{[Sr/Fe]}$ and $\mbox{[Ba/Fe]}$ across all UFD stars, with their neutron capture abundances being as much as $\sim1$ dex lower than a halo star of the same metallicity \citep{Frebel10a, Frebel10b, Frebel2014_Segue, Frebel2018_ARAA, Ji20}. When normalizing the Sr and Ba abundances to H instead of Fe to determine the overall neutron-capture enrichment level (i.e., by examining $\mbox{[Sr,Ba/H]}$ instead of $\mbox{[Sr,Ba/Fe]}$), \cite{Frebel2014_Segue} found a cleaner cut between halo stars and UFD stars, such that the latter are predominantly found with abundance levels at $\mbox{[Sr/H]} \lesssim -4$ and $\mbox{[Ba/H]} \lesssim -3.5$. Two exceptions to this rule are Reticulum II and Tucana III which show an overabundance of neutron-capture elements due to a rare $r$-process enrichment event \citep{ji16a, Roederer16, Marshall19}. Even then, two of the stars in Reticulum II still display the characteristic low neutron-capture abundances. Indeed, this chemical criterion has been crucial in evaluating the membership of candidate UFD stars \citep[e.g.,][]{chiti23}.

In this paper, we thus set out to identify metal-poor halo stars that possibly originated from small satellite galaxies, and were presumably accreted by the Milky Way long ago. We developed abundance criteria in line with chemical signatures observed among UFD stars. We then apply these criteria to our program stars and to other stars with chemical abundances from the literature. Our goal was to assemble a set of stars that could be used as local probes of the population of small satellites that assembled the Galactic halo.

We later label these probes "Small Accreted Stellar System stars" or "SASS stars." These long-lived SASS stars offer an opportunity to study, in detail, the nucleosynthesis histories of early small galaxies, their environments, and early star formation. This complements the many ongoing studies of current UFDs. Detailed comparisons of the chemical abundances of our newly identified SASS stars with those in surviving UFDs would allow for further establishing to what extent the surviving UFDs are related to the first galaxies, or at least, the earliest assembled systems \citep{Frebel2014_Segue}.

In Section~\ref{sec:observations}, we describe the initial target selection of our candidates and the spectroscopic observations. 
Sections~\ref{sec:chemanalysis} and \ref{sec:chemresults} detail the chemical abundance analysis and our results. Section \ref{sec:kinematicanalysis} presents the kinematic analysis of our program stars. In Section~\ref{sec:discussion}, we discuss our refined selection criteria for identifying SASS stars, taking into account a detailed kinematic analysis of our sample stars and considering the context of the Galactic halo's assembly. In the same section, we present possible enrichment scenarios and clues about the environments in early dwarf galaxies. We also offer future directions for identifying more SASS stars. Finally in Section~\ref{sec:conclusions}, we summarize our work.


\section{Observations and Measurements} \label{sec:observations}

\subsection{Initial target selection and existing observations}

All of our program stars were initially selected for high-resolution follow-up spectroscopy based on their brightness and low $\mbox{[Fe/H]}$ as determined from the medium-resolution spectra obtained during the Hamburg/ESO survey \citep{Wisotzki_HamburgESO, frebel06}. For this subset of stars, high-resolution spectroscopy was then obtained over several years (described further below) with the goal of carrying out a detailed chemical abundance analysis. From this sample of high-resolution spectra, we visually selected stars that showed very weak or no Ba line at 4554\,{\AA}. This preliminary selection was to establish an exploratory sample for the current study on halo stars with possible origins in long-accreted small satellite systems.

\begin{table*}
    \centering
	\caption{Observing details and heliocentric velocities}
	\begin{tabularx}{\textwidth}{Xrrrrrrrrrr}
		\hline \hline
		Star & RA & Dec & UT Dates & Slit & $t_{\text{exp}}$ & $g$ & $S/N$ & $S/N$ & $V_{\text{helio}}$ & $V_{\text{helio, Gaia}}$\\
            & [J2000]& [J2000] &  & & [\unit{s}] & [mag] & [\qty{4000}{\angstrom}] & [\qty{4500}{\angstrom}] & [\unit{km}\,\unit{s^{-1}}] & [\unit{km}\,\unit{s^{-1}}]\\
		\hline
            HE~0104$-$5300 & 01:06:51.9 & $-$52:44:10.5 & 
            Jan 6, 2013 & $1\farcs0 \times 5\farcs0$ & 1800 & 14.1 & 63 & 111 & 188.3 & 186.6 \\
            
            HE~1310$-$0536 & 13:13:31.1 & $-$05:52:12.5 & 
            Mar 11, 2014 & $0\farcs7 \times 5\farcs0$ & 3600 & 14.1 & 40 & 81 & 110.9 & 108.3\\

            HE~2155$-$2043 & 21:58:42.2 & $-$20:29:15.7 & 
            May 31, 2013 & $0\farcs7 \times 5\farcs0$ & 4339 & 13.0 & 49 & 83 & $-$94.8 & $-$90.8 \\
            
            HE~2303$-$5756 & 23:06:55.1 & $-$57:40:33.5 & 
            Jun 22, 2014 & $0\farcs7 \times 5\farcs0$ & 1200 & 13.1 & 80 & 122 & $-$85.4 & $-$92.8 \\

   		   HE~2319$-$5228 & 23:21:58.1 & $-$52:11:43.2 & 
            Jun 23, 2014 & $0\farcs7 \times 5\farcs0$ & 900 & 13.0 & 37 & 45 & 292.8 & 284.2 \\
            
		  HE~2340$-$6036 & 23:43:41.1 & $-$60:19:22.4 & 
            Sep 25, 2014 & $0\farcs7$ $\times$ $5\farcs0$ & 1200 & 12.5 & 45 & 74 & 212.4 & 212.1 \\  \hline
    \end{tabularx}
    \label{tab:observations}
\end{table*}

Since data collection occurred back in 2013 and 2014, the majority of our program stars have already been observed by other groups for a variety of purposes. Given our generally high(er) data quality, we nevertheless decided to (re-)analyze these stars to obtain our independent measurements. 

HE~0104$-$5300 was part of the large \citet{Barklem05} and \citet{Jacobson15} samples that provided detailed chemical abundances. The spectrum used in the present study is the same spectrum analyzed by \citet{Jacobson15}, allowing for a detailed abundance comparison between their work and this work.

HE~1310$-$0536 was also observed by \citet{hansen14} and noted for its low Ba and Sr abundances. Their data had a signal-to-noise ratio ($S/N$) of $\sim$40 at 4300\,{\AA} (65 at 6700\,{\AA}) whereas our spectrum has a $S/N$ of 40 at 4000\,{\AA} and $\sim$80 at 4500\,{\AA}, around where the strongest Sr and Ba absorption lines are located. 
\citet{hansen14} found [Sr/Fe] = $-$1.08 and [Ba/Fe] = $-$0.50. As we will show in Section~\ref{sec:chemresults}, we find somewhat lower values likely due to our increased $S/N$. 

Two of our carbon-rich stars, HE~2319$-$5228 and HE~2155$-$2043, were analyzed by \citet{Hansen2016} and were both found to have extremely low abundances of heavy neutron-capture elements. Our spectra are of better quality than those presented in \citet{Hansen2016}. Their spectrum of HE~2319$-$5228 had a $S/N$ of 20 at 4000\,{\AA} compared to our $\sim$40. Their spectrum of HE~2155$-$2043 was reported at 40, while we have $\sim$50. 

For HE~2319$-$5228, they report upper limits of [Ba/Fe] < $-$3 and [Sr/Fe] < $-$3. For HE~2155$-$2043, they report [Sr/Fe] = 0.2, and no value for [Ba/Fe]. 

HE~2155$-$2043 was also analyzed by \citet{Purandardas}, but no $S/N$ was reported of the spectra taken from the Subaru Telescope archive. They reported  [Sr/Fe] = $-$0.13 and [Ba/Fe] = $-$1.51. Our spectrum has a relatively high $S/N$ of $\sim$50 at 4000\,{\AA} and $\sim$80 at 4500\,{\AA}. In Section~\ref{sec:chemresults}, we will compare abundances in more detail.

HE~2340$-$6036 was included in the $r$-process Alliance's \citet{Holmbeck20} data release. A $S/N$ of 22 was reported at 4129\,{\AA} of their snapshot spectrum, compared to our $S/N$ of 45 at 4000\,{\AA} and 74 at 4500\,{\AA}. They reported weak Sr and Ba lines, finding [Sr/Fe] of $-$1.89 and [Ba/Fe] of $-$1.34. We will report somewhat higher abundances in Section~\ref{sec:chemresults}. 

Finally, we present the first detailed chemical abundance analysis of HE~2303$-$5756.

\begin{table*}
\tiny
    \centering
	\caption{Equivalent widths measurements and line abundances of our program stars}
	\label{tab:equivalentwidths}
	\begin{tabular}{lrrrrrrrrrrrrrrrrrrrrr}
		\hline \hline
            \multicolumn{4}{c}{} & 
            \multicolumn{2}{c}{HE~0104$-$5756} & \multicolumn{1}{c}{} &\multicolumn{2}{c}{HE~1310$-$0536} & \multicolumn{1}{c}{} &
            \multicolumn{2}{c}{HE 2155$-$2043} & \multicolumn{1}{c}{} & \multicolumn{2}{c}{HE 2303$-$5756} & \multicolumn{1}{c}{} &
            \multicolumn{2}{c}{HE 2319$-$5228} & \multicolumn{1}{c}{} & \multicolumn{2}{c}{HE 2340$-$6036}  \\
\cline{5-6} \cline{8-9} \cline{11-12} \cline{14-15} \cline{17-18} \cline{20-21} 
   	Species & $\lambda$ & $\chi$ & $\log gf$ & 
            EW  & $\log \epsilon$(X)  &&
            EW  & $\log \epsilon$(X)  &&
            EW  & $\log \epsilon$(X)  &&
            EW  & $\log \epsilon$(X)  &&
            EW  & $\log \epsilon$(X)  &&
            EW  & $\log \epsilon$(X)  \\
                 & [\unit{\angstrom}]& [\unit{eV}]   & [dex] &  
            [\unit{m\angstrom}] & [\unit{dex}] &&
            [\unit{m\angstrom}] & [\unit{dex}] &&
            [\unit{m\angstrom}] & [\unit{dex}] &&
            [\unit{m\angstrom}] & [\unit{dex}] &&
            [\unit{m\angstrom}] & [\unit{dex}] &&
            [\unit{m\angstrom}] & [\unit{dex}] \\	\hline
            CH&4314.00&...&...&...&4.80&&...&6.29&&...&6.00&&...&6.30&&...&6.27&&...&4.26&\\
CH&4323.00&...&...&...&4.74&&...&6.21&&...&5.99&&...&5.93&&...&6.26&&...&4.28&\\
\ion{Na}{i}&5682.63&2.10&$-$0.71&...&...&&...&...&&...&...&&...&...&&28.1&4.78&&...&...&\\
\ion{Na}{i}&5688.20&2.10&$-$0.41&...&...&&...&...&&...&...&&...&...&&33.2&4.58&&...&...&\\
\ion{Na}{i}&5889.95&0.00&+0.11&113.0&2.84&&70.4&2.20&&139.6&3.52&&45.2&3.07&&265.8&4.51&&143.4&3.19&\\
\ion{Na}{i}&5895.92&0.00&$-$0.19&107.6&3.04&&48.4&2.14&&106.4&3.22&&22.7&2.93&&242.4&4.69&&127.1&3.23&\\
\ion{Mg}{i}&4057.51&4.35&$-$0.90&11.2&4.56&&...&...&&...&...&&...&...&&...&...&&14.0&4.61&\\
\ion{Mg}{i}&4167.27&4.35&$-$0.74&18.7&4.65&&...&...&&...&...&&6.9&4.88&&...&...&&22.2&4.68&\\
\ion{Mg}{i}&4702.99&4.33&$-$0.44&24.4&4.44&&10.4&4.00&&35.1&4.76&&10.5&4.74&&58.9&5.03&&36.5&4.60&\\
\ion{Mg}{i}&5172.68&2.71&$-$0.36&141.1&4.38&&88.1&3.60&&147.2&4.70&&89.5&4.81&&176.0&4.83&&160.4&4.45&\\
\ion{Mg}{i}&5183.60&2.72&$-$0.17&160.1&4.46&&105.4&3.76&&163.0&4.75&&104.0&4.90&&187.4&4.75&&177.2&4.45&\\
\ion{Al}{i}&3944.00&0.00&$-$0.64&...&2.60&&...&2.08&&...&...&&...&2.76&&...&...&&...&2.34&\\
\ion{Al}{i}&3961.52&0.01&$-$0.34&...&2.13&&...&1.31&&...&...&&...&2.71&&...&2.82&&...&2.30&\\
\ion{Si}{i}&3905.52&1.91&$-$1.09&...&4.54&&...&...&&...&...&&...&4.67&&...&4.92&&...&4.61&\\
\ion{Si}{i}&4102.94&1.91&$-$3.14&...&...&&...&...&&...&...&&...&...&&...&4.23&&...&4.98&\\
        \hline
  \multicolumn{16}{l}{This table is available in its entirety in the online journal.}
	\end{tabular}
\end{table*}

\begin{table*}
	\caption{Photometry and astrometry for our program stars}
	\label{tab:photometry}
	\begin{tabular}{llrrrrrr} \hline \hline
		Star & \textit{Gaia} Source ID & J & K & V & $A_V$ & $E(B-V)$ & parallax (\unit{mas}) \\ \hline
HE~0104$-$5300 & 4927175937828177280 & $11.91\pm0.03$ & $11.29\pm0.02$ & $13.63\pm0.03$ & 0.0326 & $0.0108\pm0.0004$& $0.0999\pm0.0112$ \\

HE~1310$-$0536 & 3635533208672382592 & $12.53\pm0.02$ & $12.05\pm0.02$ & $14.38\pm0.02$ & 0.1149 & $0.0368\pm0.0016$& $0.0118\pm0.0232$ \\ 

HE~2155$-$2043 & 6823219517280229120 & $11.57\pm0.02$ & $11.02\pm0.02$ & $13.23\pm0.01$ & 0.0799 & $0.0257\pm0.0006$& $0.1240\pm0.0301$ \\

HE~2303$-$5756 & 6493230587155263360 & $12.23\pm0.02$ & $11.91\pm0.02$ & $13.24\pm0.02$ & 0.0455 & $0.0147\pm0.0005$& $0.9229\pm0.0166$ \\

HE~2319$-$5228 & 6501398446721935744 & $11.55\pm0.02$ & $10.96\pm0.02$ & $13.25\pm0.02$ & 0.0277 & $0.0087\pm0.0005$ & $0.1616\pm0.0158$ \\

HE~2340$-$6036 & 6488378098744387968 & $11.05\pm0.02$ & $10.40\pm0.02$ & $12.83\pm0.01$ & 0.0363 & $0.0119\pm0.0002$& $0.1666\pm0.0126$ \\
\hline
\end{tabular}
\end{table*}

\subsection{High-resolution spectroscopy}

All of our stars were observed at Las Campanas Observatory in January and May of 2013, as well as March, June, and September of 2014 using the Magellan Inamori Kyocera Echelle (MIKE) instrument \citep{Bernstein03} on the Magellan-Clay telescope. 
Data for all program stars except HE 0104$-$5300 was taken with a slit size of $0\farcs7 \times 5\farcs0$, resulting in a resolving power of $\sim$28,000 in the red wavelength and $\sim$35,000 in the blue. Data for HE 0104$-$5300 was obtained with a $1\farcs0 \times 5\farcs0$ slit, yielding a resolving power of $\sim$22,000 and $\sim$28,000 for the red and blue spectra, respectively. Exposure times ranged from 900 to roughly 4500 seconds.

\begin{figure*}
    \raggedleft
        \includegraphics[width=1\linewidth]{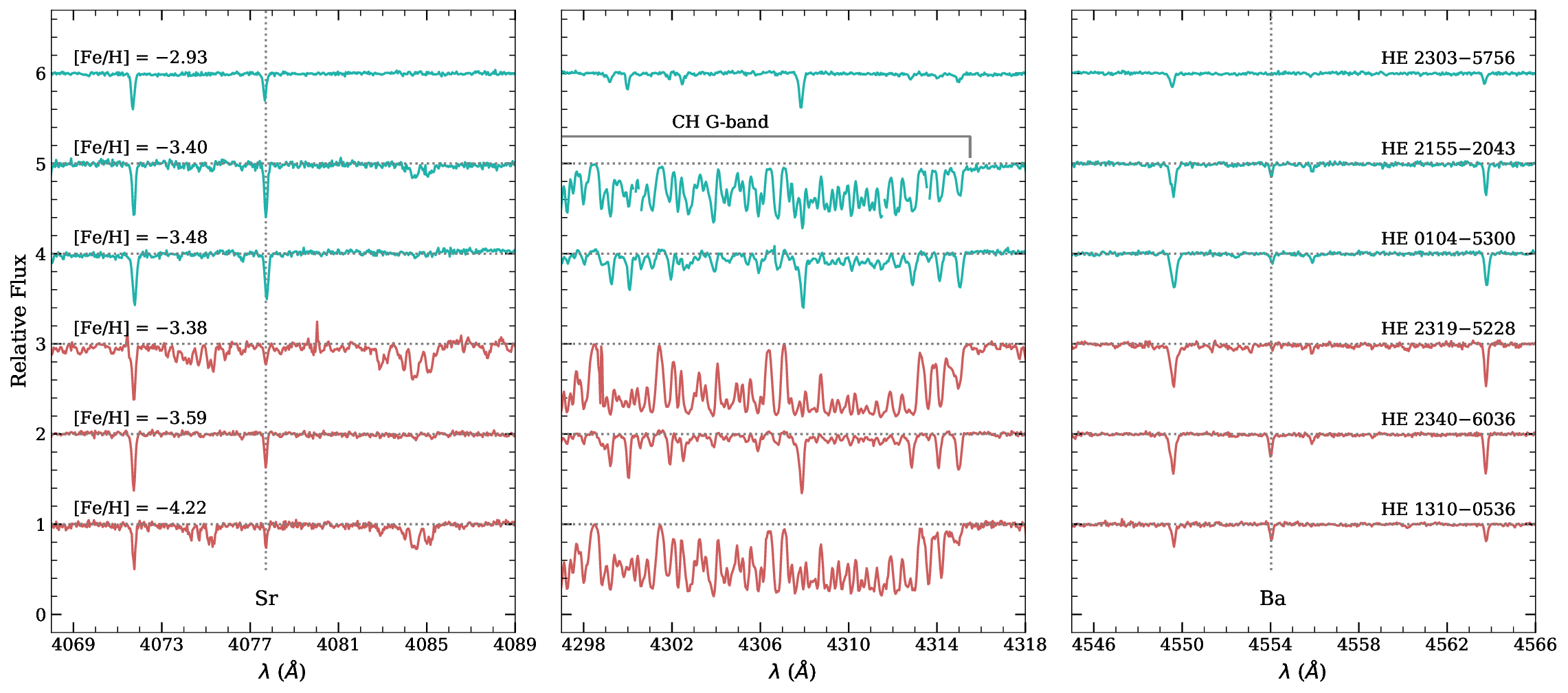}
\caption{Spectra of our program stars around the Sr line at 4077\,{\AA}, the carbon G-band around 4300\,{\AA}, and the Ba line at 4554\,{\AA}. 
The top three spectra (teal)
are halo stars from our sample (HE 2303$-$5756, HE 2155$-$2043, HE 0104$-$5300), whereas the bottom three spectra (red) depict stars with putative origins
in long-accreted small satellite systems (HE 2319$-$5228, HE 2340$-$6036, HE 1310$-$0536). Spectra were shifted vertically to aid visual representation. Note
that HE 2303$-$5756 (teal) and HE 2319$-$5228 (red) appear to be in binary systems.}
        \label{fig:spectralregions}
\end{figure*}

We used the CarPy MIKE pipeline (the version available at the time) to reduce all spectra \citep{Kelson03}. Using the SMH custom analysis software \citep{Casey2014}, we normalized and stitched all blue and red echelle orders together before Doppler correcting the spectra by cross-correlation. We employed HD122563 as a template using the Ca triplet lines around 8500\,{\AA}. Heliocentric velocities were determined using standard Python packages. Observational details and radial velocity measurements are listed in Table~\ref{tab:observations}. Figure~\ref{fig:spectralregions} shows three selected spectral regions of the final spectra of all our stars: around the Sr line at 4077\,{\AA}, the CH G-band around 4300\,{\AA}, and the Ba line at 4554\,{\AA}.

We also list \textit{Gaia} radial velocities in Table~\ref{tab:observations} for comparison with our derived values. The measurements generally agree within a few km\,s$^{-1}$. The largest differences are found for HE~2319$-$5228 and HE~2303$-$5756 for which the radial velocities differ by nearly 10 km\,s$^{-1}$, suggesting that these stars are part of binary systems.
Searching for additional radial velocity measurements in the SIMBAD database reveals further support for their binarity. We note that neither of the two stars appears to be a double-lined spectroscopic binary so our spectra are not affected by any light from a presumably faint companion. 

HE~2319$-$5228 is very carbon-enhanced which could be due to a mass transfer event from its previously more massive companion. Since it is not $s$-process-enriched, as is common for carbon-rich stars in binary systems, it may have had a relatively massive erstwhile AGB companion. An alternative enrichment scenario that could explain the observed abundances is a massive, rapidly rotating "spinstar" that produced large amounts of carbon and then enriched the natal gas cloud of HE~2319$-$5228 \citep{chiappini06, Cescutti_2013_spinstars, Meynet06, Hirschi06}. If such spinstars were present among later generations of stars (not just among Pop\,III stars), then HE~2319$-$5228 may have had a spinstar companion that transferred carbon-rich mass into it.

\section{Chemical Abundance Analysis} \label{sec:chemanalysis}

\subsection{Line measurements}

To determine the chemical abundances of our stars, we utilize a line list from linemake\footnote{Available at \url{https://github.com/vmplacco/linemake}} \citep{Placco21} that contains absorption lines of various elements. We measure the strength of clean spectral lines by fitting Gaussian profiles to each absorption feature, after which we obtain their equivalent widths (EWs). Only lines between $\qty{3500}{\angstrom}$ and $\qty{7000}{\angstrom}$ were used. Below $\qty{4000}{\angstrom}$ the $S/N$ becomes lower, making good line measurements challenging. To ensure that EWs are on the linear part of the curve of growth, only lines with reduced EWs less than $-4.5$ were used. Some exceptions were made for elements with few lines, such as Na, but line abundances were always individually checked for consistency.

EW measurements are not useful for spectral features that are blended,  lines with hyperfine structure, and molecular bands (e.g., CH). In those cases, we used the spectrum synthesis approach to obtain the line abundance where we fit a synthetic spectrum with a known abundance to the observed spectrum. Spectrum synthesis was performed for features of C, Si, Al, Sc, Mn, Co, Sr,  Ba, and Eu. We also determined $3\sigma$ upper limits on the abundances of any pertinent, undetected lines. Table~\ref{tab:equivalentwidths} lists all our measured EWs (where available) and their corresponding abundances.

\subsection{Stellar Parameters}
\label{sec:stellarparams}

We determine photometric stellar parameters using parallaxes and proper motion data from \textit{Gaia} (where available) and our line measurements of \ion{Fe}{i} and \ion{Fe}{ii} lines. First, we determine the effective temperature ($T_{\text{eff}}$) using the $V-K$ color-temperature-metallicity formula presented in \cite{Casagrande2010}. Using the United States Naval Observatory UCAC4 Catalog\footnote{Accessible via \href{https://irsa.ipac.caltech.edu/applications/Gator/index.html}{NASA/IPAC Infrared Science Archive}} \citep{Zacharias13_UCAC4}, we utilize J and K magnitudes from 2MASS (Two Micron All-Sky Survey) and V magnitudes from APASS (AAVSO Photometric All-Sky Survey). To correct for dust reddening, we use the extinction ($A_V$) and reddening $E(B-V)$ values from the dust maps by \citet{Schlafly&Finkbeiner2011}. Values are listed in Table~\ref{tab:photometry}. As necessary, we iterate on the temperature determination with our simultaneous metallicity determination.

We calculate the surface gravity ($\log g$) by minimizing the trend between \ion{Fe}{I} and \ion{Fe}{II} lines. The values are listed in Table~\ref{tab:stellarparams}. We then test the robustness of our results by re-deriving $\log g$ using a different method, specifically the one used in \citet{mardini19}. This method uses dereddened $g$ magnitudes from  APASS (compiled on the UCAC4 catalog), parallaxes $\pi$ from \textit{Gaia} DR3 \citep{GaiaDR3}, and fundamental equations with respect to the Sun. The surface gravity values derived with both methods agree within $1\sigma$.

Generally, our stellar parameters agree well with those obtained by other studies (e.g., \citealt{Jacobson15}). For one star, HE~1310$-$0536, our $T_{\text{eff}}$ of $\sim$4700\,K is lower than that derived by \cite{hansen14} ($T_{\text{eff}}$ = 5000\,K). However, a visual comparison of this star's H~$\beta$ absorption line profile with those of stars with known temperatures — namely CS22892-52 ($T_{\text{eff}}$ = 4620\,K) and HE~1523$-$0901 ($T_{\text{eff}}$ = 4370\,K) — well supports $T_{\text{eff}}$ < 5000\,K.

For our abundance calculations, we adopt 1D plane-parallel model atmospheres with $\alpha$-enhancement (generated using the values listed in Table~\ref{tab:stellarparams}) from \citet{CastelliKurucz2004}. We used the latest version of the MOOG radiative transfer code \citep{Sneden1973, Sobeck2011} and assumed local thermodynamic equilibrium (LTE). This MOOG version includes scattering effects. We then obtain the Fe abundance for each star using their respective effective temperatures and surface gravities, as determined above and including any iterations. 

In the process of determining the Fe abundances, we also derived the microturbulence ($v_{\text{mic}}$) for each star by forcing no trend between \ion{Fe}{i} line abundance and reduced equivalent widths. All our abundance ratios ([X/Fe] and [X/H]) are calculated relative to the solar abundances determined by \citet{Asplund2009}. Our final and adopted stellar parameters are given in Table~\ref{tab:stellarparams}.

\begin{table}
\centering
	\caption{Stellar parameters}
	\label{tab:stellarparams}
	\begin{tabular}{lrrrr} \hline \hline
	Star & $T_{\text{eff}}$ & $\log g$   &   [Fe/H]     & $v_{\text{mic}}$  \\
             & [\unit{K}]  & [\unit{cgs}]& [\unit{dex}] & [\unit{km\,s^{-1}}]\\	\hline
	HE 0104$-$5300 & 4791 & 1.82 & $-3.48 \pm 0.15$ & 1.95\\
        HE 1310$-$0536 & 4668 & 0.57 & $-4.22 \pm 0.14$ & 1.84\\
        HE 2155$-$2043 & 4994 & 1.82 & $-3.40 \pm 0.17$ & 2.02\\
        HE 2303$-$5756 & 6349 & 3.60 & $-2.93 \pm 0.10$ & 1.53\\
        HE 2319$-$5228 & 4836 & 2.09 & $-3.38 \pm 0.12$ & 1.70\\
	HE 2340$-$6036 & 4689 & 1.93 & $-3.59 \pm 0.17$ & 1.91\\
        	\hline
	\end{tabular}
\end{table}


\section{Chemical Abundance Results}
\label{sec:chemresults}

\begin{table*}\centering
    \caption{Chemical abundances of our program stars}
    \label{tab:chemresults}
    \begin{tabularx}{\textwidth}{Xlrrrr|Xlrrrr}
    \hline \hline
    Species & $N$ & $\log\epsilon(X)$ & stderr & $\mbox{[X/H]}$ & $\mbox{[X/Fe]}$ & 
    Species & $N$ & $\log\epsilon(X)$ & stderr & $\mbox{[X/H]}$ & $\mbox{[X/Fe]}$ \\ 
    \hline
    \multicolumn{6}{c}{\textbf{HE~0104$-$5300}} & 
    \multicolumn{6}{c}{\textbf{HE~1310$-$0536}} \\
    \hline
    C (CH) & 2 & 4.77 & 0.10 & $-$3.66 & $-$0.17 & C (CH) & 2 & 6.25 & 0.10 & $-$2.18 & 2.04\\
    C (CH)$_{\rm corr}$ & ... & ... & ... & ... & $-$0.11 & C (CH)$_{\rm corr}$ & ... & ... & ... & ... & 2.67 \\
    \ion{Na}{i} & 2 & 2.94 & 0.13 & $-$3.30 & 0.18 & \ion{Na}{i} & 2 & 2.17 & 0.05 & $-$4.07 & 0.15 \\
    \ion{Mg}{i} & 5 & 4.50 & 0.05 & $-$3.10 & 0.38 & \ion{Mg}{i} & 6 & 3.84 & 0.06 & $-$3.76 & 0.46  \\
    \ion{Al}{i} & 2 & 2.37 & 0.42 & $-$4.08 & $-$0.60 & \ion{Al}{i} & 2 & 1.70 & 0.51 & $-$4.75 & $-$0.53 \\
    \ion{Si}{i} & 2 & 4.67 & 0.16 & $-$2.84 & 0.64 & \ion{Si}{i} & 1 & 3.68 & 0.15 & $-$3.83 & 0.39 \\
    \ion{Ca}{i} & 14 & 3.22 & 0.05 & $-$3.12 & 0.36 & \ion{Ca}{i} & 6 & 2.78 & 0.05 & $-$3.56 & 0.66 \\ 
    \ion{Sc}{ii} & 6 & $-$0.29 & 0.05 & $-$3.44 & 0.05 & \ion{Sc}{ii} & 1 & $-$1.19 & 0.15 & $-$4.34 & $-$0.12 \\ 
    \ion{Ti}{i} & 12 & 1.74 & 0.05 & $-$3.21 & 0.27 & \ion{Ti}{i} & 3 & 1.19 & 0.24 & $-$3.76 & 0.47 \\
    \ion{Ti}{ii} & 21 & 1.91 & 0.05 & $-$3.04 & 0.45 & \ion{Ti}{ii} & 14 & 1.28 & 0.05 & $-$3.67 & 0.55 \\
    \ion{Cr}{i} & 5 & 1.89 & 0.13 & $-$3.75 & $-$0.26 & \ion{Cr}{i} & 1 & 1.07 & 0.15 & $-$4.57 & $-$0.35 \\
    \ion{Mn}{i} & 1 & 1.07 & 0.15 & $-$4.36 & $-$0.88 & \ion{Mn}{i} & 1 & 0.36 & 0.15 & $-$5.07 & $-$0.85 \\
    \ion{Fe}{i} & 116 & 4.02 & 0.05 & $-$3.48 & 0.00 & \ion{Fe}{i} & 43 & 3.28 & 0.10 & $-$4.22 & 0.00 \\
    \ion{Fe}{ii} & 13 & 4.27 & 0.05 & $-$3.23 & 0.25 & \ion{Fe}{ii} & 2 & 3.25 & 0.33 & $-$4.25 & $-$0.03 \\
    \ion{Co}{i} & 3 & 1.82 & 0.05 & $-$3.17 & 0.31 & \ion{Co}{i} & 2 & 0.86 & 0.05 & $-$4.13 & 0.09 \\
    \ion{Ni}{i} & 6 & 2.92 & 0.12 & $-$3.30 & 0.18 & \ion{Ni}{i} & 6 & 1.95& 0.07 & $-$4.27 & $-$0.05 \\
    \ion{Zn}{i} & 1 & 1.64 & 0.15 & $-$2.92 & 0.56 & \ion{Zn}{i} & 1 & $<$ 0.57 & ... & $<-$3.98 & $<$ 0.24 \\
    \ion{Sr}{ii} & 2 & $-$1.02 & 0.05 & $-$3.89 & $-$0.40 & \ion{Sr}{ii} & 1 &$-$2.99 & 0.15 &$-$5.86&$-$1.64 \\
    \ion{Ba}{ii} & 2 & $-$2.71 & 0.05 & $-$4.89 & $-$1.41 & \ion{Ba}{ii} & 2 & $-$2.92 & 0.15 & $-$5.10 & $-$0.88 \\
    \ion{Eu}{ii} & 1 & $<-$2.62 & ... & $<-$3.14 & $<$ 0.34 & \ion{Eu}{ii} & 1 & $<-$3.20 & ... & $<-$3.72 & $<$ 0.50 \\
    \hline
    \multicolumn{6}{c}{\textbf{HE~2155$-$2043}} & 
    \multicolumn{6}{c}{\textbf{HE~2303$-$5756}} \\
    \hline
    C (CH) & 2 & 6.00 & 0.10 & $-$2.43 & 0.97 & C (CH) & 2 & 6.11 & 0.35 & $-$2.32 & 0.61 \\
    C (CH)$_{\rm corr}$ & ... & ... & ... & ... & 1.07 & C (CH)$_{\rm corr}$ & ... & ... & ... & ... & 0.61 \\
    \ion{Na}{i} & 2 & 3.37 & 0.21 & $-$2.87 & 0.54 & \ion{Na}{i} & 2 & 3.00 & 0.09 & $-$3.24 & $-$0.31 \\ 
    \ion{Mg}{i} & 8 & 4.72 & 0.05 & $-$2.88 & 0.53 & \ion{Mg}{i} & 7 & 4.82 & 0.05 & $-$2.78 & 0.15 \\
    \ion{Al}{i} & 1 & 2.29 & 0.10 & $-$4.16 & $-$0.76 & \ion{Al}{i} & 2 & 2.74 & 0.05 & $-$3.71 & $-$0.79 \\
    \ion{Si}{i} & 1 & 4.06 & 0.20 & $-$3.45 & $-$0.04 & \ion{Si}{i} & 1 & 4.67 & 0.15 & $-$2.84 & 0.09 \\
    \ion{Ca}{i} & 11 & 3.32 & 0.05 & $-$3.02 & 0.38 & \ion{Ca}{i} & 9 & 3.76 & 0.05 & $-$2.58 & 0.34 \\
    \ion{Sc}{ii} & 5 & $-$0.22 & 0.05 & $-$3.37 & 0.03 & \ion{Sc}{ii} & 4 & 0.29 & 0.05 & $-$2.86 & 0.06 \\
    \ion{Ti}{i} & 8 & 1.93 & 0.08 & $-$3.02 & 0.38 & \ion{Ti}{i} & 3 & 2.62 & 0.05 & $-$2.33 & 0.60 \\
    \ion{Ti}{ii} & 17 & 1.70 & 0.05 & $-$3.25 & 0.15 & \ion{Ti}{ii} & 14 & 2.23 & 0.05 & $-$2.72 & 0.21 \\
    \ion{Cr}{i} & 3 & 1.93 & 0.05 & $-$3.71 & $-$0.30 & \ion{Cr}{i} & 4 & 2.62 & 0.05 & $-$3.02 & $-$0.09 \\
    \ion{Mn}{i} & 1 & 1.29 & 0.20 & $-$4.14 & $-$0.73 & \ion{Mn}{i} & 1 & 1.94 & 0.15 & $-$3.49 & $-$0.57 \\
    \ion{Fe}{i} & 110 & 4.10 & 0.05 & $-$3.40 & 0.00 & \ion{Fe}{i} & 67 & 4.57 & 0.05 & $-$2.93 & 0.00 \\
    \ion{Fe}{ii} & 10 & 4.18 & 0.05 & $-$3.32 & 0.08 & \ion{Fe}{ii} & 5 & 4.53 & 0.05 & $-$2.97 & $-$0.04 \\
    \ion{Co}{i} & 7 & 2.05 & 0.05 & $-$2.94 & 0.47 & \ion{Co}{i} & 6 & 2.47 & 0.05 & $-$2.52 & 0.41 \\
    \ion{Ni}{i} & 11 & 2.92 & 0.05 & $-$3.30 & 0.10 & \ion{Ni}{i} & 13 & 3.52 & 0.05 & $-$2.70 & 0.23 \\
    \ion{Zn}{i} & 2 & 1.77 & 0.21 & $-$2.79 & 0.61 & \ion{Zn}{i} & 1 & $<$ 2.16 & ... & $<-$2.40 & $<$ 0.53 \\
    \ion{Sr}{ii} & 2 & $-$0.76 & 0.06 & $-$3.63 & $-$0.23 & \ion{Sr}{ii} & 2 & $-$0.47 & 0.05 & $-$3.34 & $-$0.41 \\
    \ion{Ba}{ii} & 1& $-$2.38 & 0.08 & $-$4.56 & $-$1.16 & \ion{Ba}{ii} & 1 & $<-$1.75 & ... & $<-$3.93 & $<-$1.00 \\
    \ion{Eu}{ii} & 1 & $<-$2.50 & ... & $<-$3.02 & $<$ 0.38 & \ion{Eu}{ii} & 1 & $<-$1.42 & ... & $<-$1.94 & $<$ 0.99 \\
    \hline
    \multicolumn{6}{c}{\textbf{HE 2319$-$5228}} & 
    \multicolumn{6}{c}{\textbf{HE 2340$-$6036}} \\
    \hline
    C (CH) & 2 & 6.27 & 0.10 & $-$2.16 & 1.21 & C (CH) & 2 & 4.27 & 0.15 & $-$4.16 & $-$0.56\\
    C (CH)$_{\rm corr}$ & ... & ... & ... & ... & 1.22 & C (CH)$_{\rm corr}$ & ... & ... & ... & ... & $-$0.55 \\
    \ion{Na}{i} & 4 & 4.64 & 0.07 & $-$1.60 & 1.78 & \ion{Na}{i} & 2 & 3.21 & 0.05 & $-$3.03 & 0.56 \\ 
    \ion{Mg}{i} & 3 & 4.87 & 0.10 & $-$2.73 & 0.65 & \ion{Mg}{i} & 5 & 4.56 & 0.05 & $-$3.04 & 0.55 \\ 
    \ion{Al}{i} & 1 & 2.82 & 0.20 & $-$3.63 & $-$0.25 & \ion{Al}{i} & 2 & 2.32 & 0.05 & $-$4.13 & $-$0.53 \\ 
    \ion{Si}{i} & 2 & 5.07 & 0.05 & $-$2.44 & 0.94 & \ion{Si}{i} & 2 & 4.79 & 0.23 & $-$2.72 & 0.88 \\ 
    \ion{Ca}{i} & 17 & 3.96 & 0.05 & $-$2.38 & 0.99 & \ion{Ca}{i} & 13 & 3.40 & 0.05 & $-$2.94 & 0.65 \\ 
    \ion{Sc}{ii} & 6 & 0.06 & 0.05 & $-$3.09 & 0.29 & \ion{Sc}{ii} & 6 & $-$0.21 & 0.05 & $-$3.36 & 0.24 \\ 
    \ion{Ti}{i} & 7 & 1.99 & 0.05 & $-$2.96 & 0.42 & \ion{Ti}{i} & 9 & 1.58 & 0.05 & $-$3.37 & 0.22 \\ 
    \ion{Ti}{ii} & 17 & 2.34 & 0.05 & $-$2.61 & 0.77 & \ion{Ti}{ii} & 21 & 1.95 & 0.05 & $-$3.00 & 0.59 \\
    \ion{Cr}{i} & 3 & 2.11 & 0.17 & $-$3.53 & $-$0.15 & \ion{Cr}{i} & 4 & 1.54 & 0.08 & $-$4.10 & $-$0.50 \\
    \ion{Mn}{i} & 2 & 1.40 & 0.15 & $-$4.03 & $-$0.66 & \ion{Mn}{i} & 1 & 0.70 & 0.15 & $-$4.73 & $-$1.14 \\
    \ion{Fe}{i} & 59 & 4.12 & 0.05 & $-$3.38 & 0.00 & \ion{Fe}{i} & 86 & 3.91 & 0.05 & $-$3.59 & 0.00 \\
    \ion{Fe}{ii} & 6 & 4.29 & 0.05 & $-$3.21 & 0.16 & \ion{Fe}{ii} & 8 & 4.23 & 0.07 & $-$3.27 & 0.32 \\
    \ion{Co}{i} & 3 & 1.66 & 0.05 & $-$3.33 & 0.05 & \ion{Co}{i} & 3 & 1.51 & 0.13 & $-$3.48 & 0.11 \\
    \ion{Ni}{i} & 1 & 2.71 & 0.15 & $-$3.51 & $-$0.13 & \ion{Ni}{i} & 2 & 2.41 & 0.06 & $-$3.81 & $-$0.22 \\
    \ion{Zn}{i} & 1 & 1.73 & 0.25 & $-$2.83 & 0.55 & \ion{Zn}{i} & 1 & $<$ 1.57 & ... & $<-$2.99 & $<$ 0.60 \\
    \ion{Sr}{ii} & 1 & $-$2.31 & 0.15 & $-$5.18 & $-$1.81 & \ion{Sr}{ii} & 2 & $-$2.13 & 0.15 & $-$5.00 & $-$1.41 \\
    \ion{Ba}{ii} & 1 & $-$3.20 & 0.15 &$-$5.38 & $-$2.01 & \ion{Ba}{ii} & 3 & $-$2.26 & 0.15 & $-$4.44 & $-$0.85 \\
    \ion{Eu}{ii} & 1 & ... & ... & ... & ... & \ion{Eu}{ii} & 1 & $<-$2.77 & ... & $<-$3.29 & $<$ 0.31 \\
    \hline
   \multicolumn{12}{l}{For Zn, upper limits were determined from the line at 4810\,{\AA}. For Eu, they were derived from the line at 4129\,{\AA}.}
\end{tabularx} 
\end{table*}

\begin{figure*}
        \includegraphics[width=\linewidth]{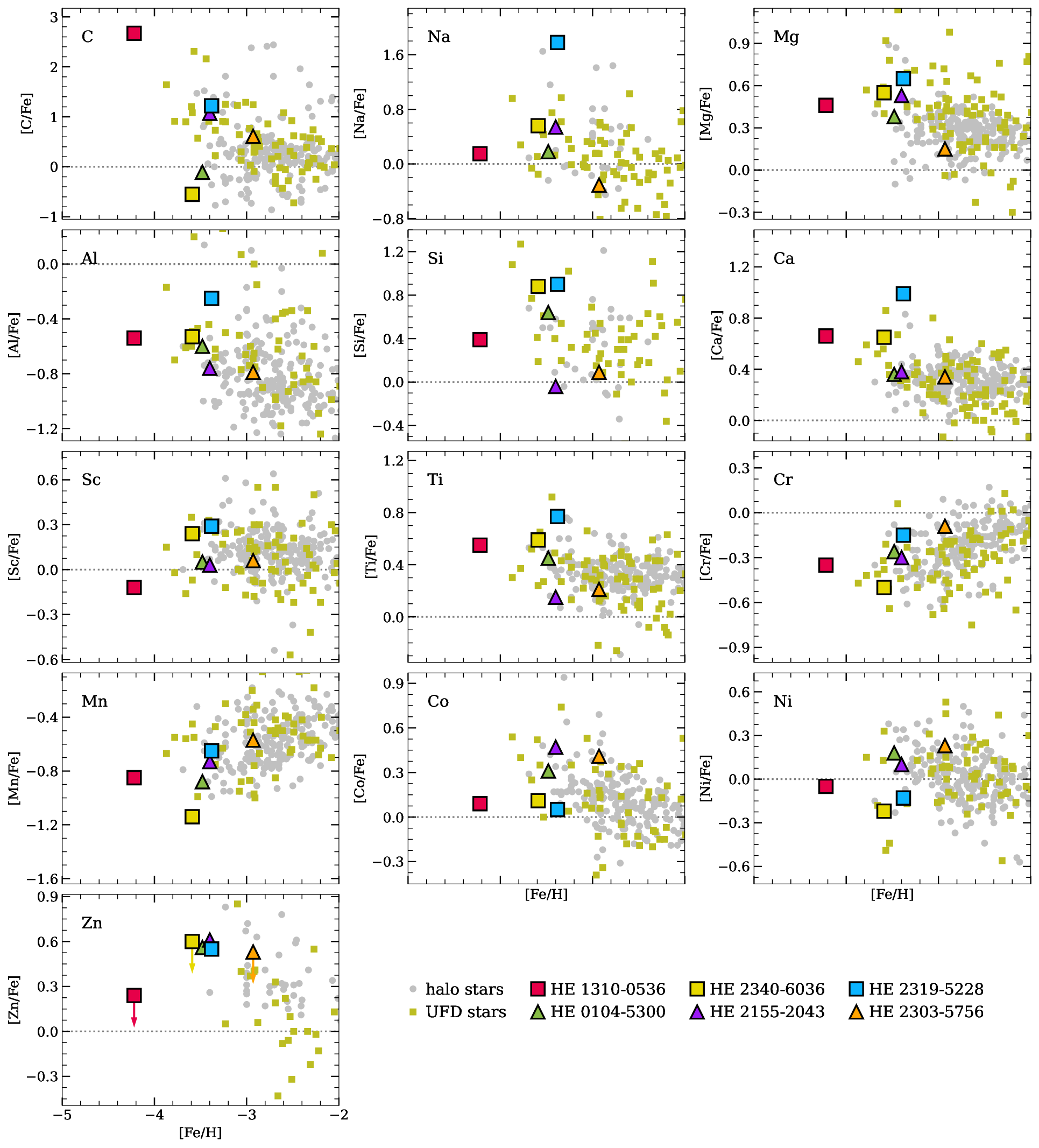} 
        \vspace{-0.4cm}
         \caption{ \label{fig:light abundances} Light element abundances for our program stars (squares and triangles with black outlines), UFD stars (olive squares; see Section~\ref{sec:chemresults} for references), and halo stars (gray points; \citealt{Cayrel04, Barklem05, Yong13a}). The abundances of the stars denoted by triangles (HE 0104$-$5300, HE 2155$-$2043, HE 2303$-$5756) generally follow the abundance trends found in halo stars. Meanwhile, the stars denoted by squares (HE 1310$-$0536, HE 2340$-$6036, HE 2319$-$5228) show more variation and deviations from the general halo trend.}
\end{figure*}

In Table~\ref{tab:chemresults}, we present the results of our detailed chemical abundance analysis of 17 elements in our sample stars. In Figure~\ref{fig:light abundances}, we show our sample stars in comparison with a general set of metal-poor halo stars \citep{Cayrel04, Barklem05, Yong13a} and with metal-poor stars in various UFDs\footnote{Compilation by AP Ji available on \href{https://github.com/alexji/alexmods/blob/master/alexmods/data/abundance_tables/dwarf_lit_all.tab}{GitHub}} \citep{Koch08, Simon2010_LeoIV, Francois16, frebel10, Norris10b, ji16a, Koch13, Gilmore13, Ishigaki14, Frebel2014_Segue, Ji2016_RetIINature, Roederer16, frebel16, kirby2017_tri, Spite18, Chiti18, nagasawa2018, Ji19, Marshall19, Ji20, Hansen20, Waller22, Chiti22}. In the following, we discuss notable results and abundance trends. Further discussion of the origins of our stars and their possible relations to early, small dwarf galaxies will come in Section~\ref{sec:discussion}.

As mentioned in Section~\ref{sec:stellarparams}, we derived $\log g$ using two methods, i.e., spectroscopically versus using photometry and parallax. To err on the side of caution, we tested whether changes in stellar parameters affected the chemical abundances. We found that the abundances were not significantly altered. Since our abundance measurements are robust to changes in $\log g$, from this point onward we will only report abundances obtained using the spectroscopically-derived $\log g$.

\subsection{Carbon abundances}

The most metal-poor halo stars with $\mbox{[Fe/H]}<-2.5$ have long been known to have predominantly very large carbon abundances (e.g., \citealt{frebel05, Placco14, Hansen2016_CEMPno, Lee2017_CHalo, Yoon2018_AEGISCFe}). In general, carbon enhancement is defined as $\mbox{[C/Fe]}>0.7$ \citep{Aoki07}. As stars ascend the giant branch, carbon abundances get depleted. We correct for this fact using models provided in \citet{Placco14}. Our sample shows a large variety of [C/Fe], as measured from the carbon G-band head around 4313\,{\AA} and the feature at 4323\,{\AA} (see Figure~\ref{fig:spectralregions}). We used spectrum synthesis on each spectral feature to infer the star's carbon abundance, after which we averaged the measurements on both features to obtain the star's final carbon abundance.

Three of our program stars, HE~1310$-$0536, HE~2155$-$2043, and HE~2319$-$5228 can thus be classified as carbon-enhanced metal-poor (CEMP) stars. HE~1310$-$0536, with the lowest [Fe/H] of $-$4.22, strongly follows the halo trend having [C/Fe] = 2.7 (corrected). The other two stars have corrected abundances around [C/Fe] $\sim$ 1.2 and also support the halo trend. In contrast, two of our sample stars have very low, subsolar carbon abundances (HE~2340$-$6036 and HE~0104$-$5300) at [Fe/H] $\sim-$3.5.

\subsection{Light and iron-peak element abundance trends} \label{sec:light iron trends}

In general, our stars' chemical abundances agree well with the overall abundance trends of halo and dwarf galaxy stars. However, it is interesting to note that we do see a significant spread among our six stars. We recall in this context that these stars were selected only visually for their having a weak or no Ba line in their spectrum. This selection alone might imply a certain level of heterogeneity among light and iron-peak elements as measured in our sample but it may not warrant it either. In Figure~\ref{fig:light abundances}, it can be seen that for several elements (Al, Ca, Ti, Co, Ni), the stars denoted by squares (HE 1310$-$0536, HE 2340$-$6036, HE 2319$-$5228) differ somewhat from the stars denoted by triangles (HE 0104$-$5300, HE 2155$-$2043, HE 2303$-$5756). We chose to group these stars by shape according to their behaviour in the [Sr/Ba] versus [Ba/Fe] plot (see Figure~\ref{fig:srba} and Section~\ref{sec:heavy}). Stars denoted by squares behave like UFD stars in Figure~\ref{fig:srba} while those denoted by triangles agree with typical halo star [Sr/Ba] behavior.

Generally, the stars denoted by triangles reasonably follow the overall halo trends in their light element abundances. Meanwhile, the stars denoted by squares tend to deviate from them. For Zn, we only have two measurements, so no strong conclusions can be derived. However, the upper limit of HE~1310$-$0536, our most metal-poor star, is significantly lower than the halo trend. We thus add Zn to the list of elements for which the two star groups show different behavior.

\subsection{Heavy element abundance trends} \label{sec:heavy}

Sr was detected in all six of our sample stars. All have [Sr/H] $< -$3, and three have a remarkably low abundance of [Sr/H] $\lesssim-$5. Ba abundances were determined for all stars except HE~2303$-$5756, for which only an upper limit of [Ba/H] $< -$3.93 was found. The other five stars all had [Ba/H] $< -$4.0, with three having [Ba/H] $< -$5.0. For [Sr/H] and [Ba/H], all stars follow the general halo trend. 

\begin{figure}
\hspace{-0.5cm}
        \includegraphics[width=1.05\linewidth]{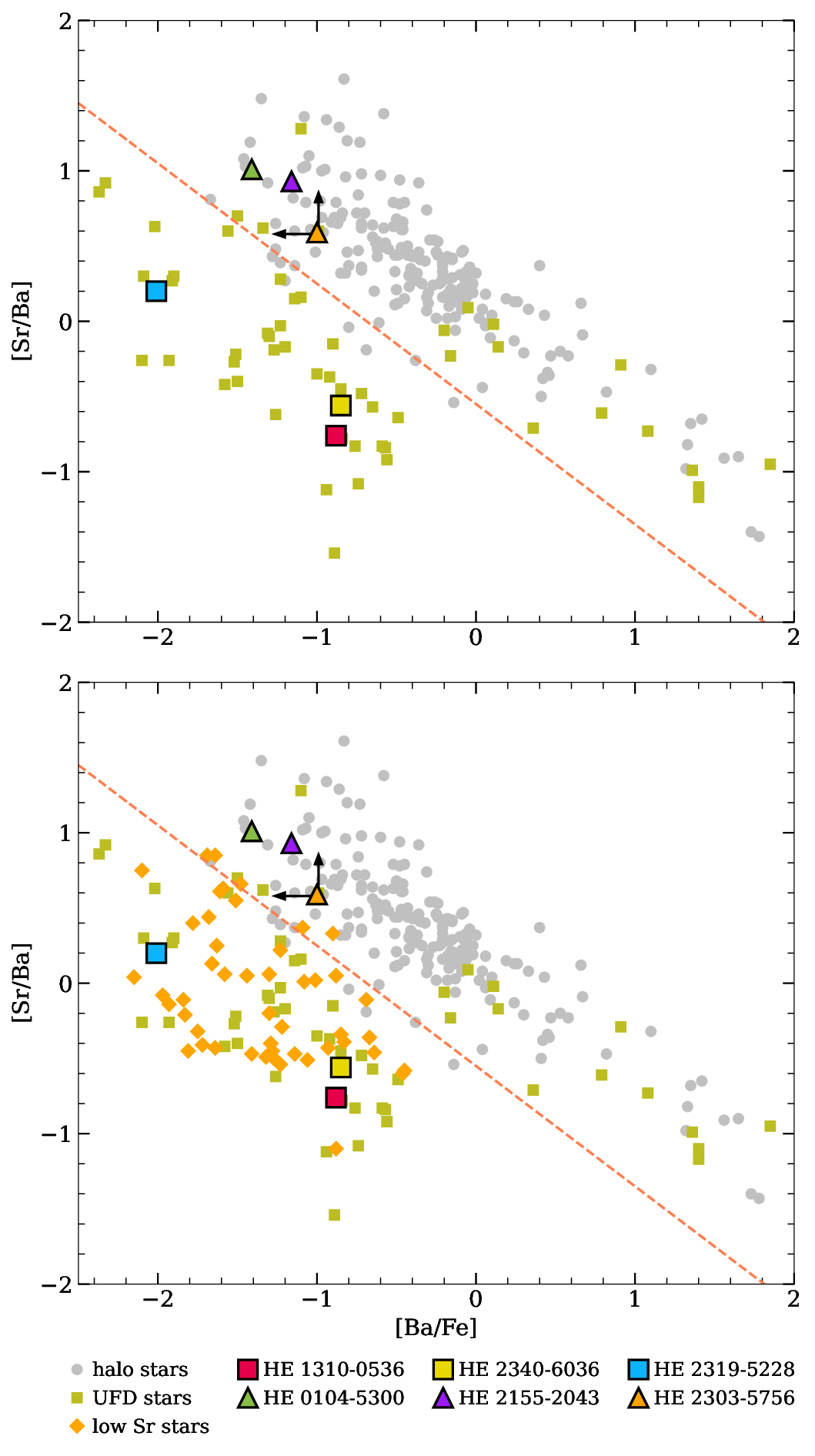}
        \caption{Top: [Sr/Ba] vs [Ba/Fe] for our sample stars in comparison with halo stars (gray points) and UFD stars (olive squares). The diagonal line guides the eye and separates halo stars from UFD stars. The UFD stars towards the bottom right corner that appear to extend the main halo trend are the highly $r$-process enhanced stars in Reticulum\,II \citep{Ji2016_RetIINature} as this is the region occupied by strongly enhanced $r$-process stars. Bottom: Same as above but with additional halo stars (orange diamonds) that have been selected with a Sr cut of [Sr/H] $<-4.5$. They appear to overlap with the region covered by UFD stars.}
        \label{fig:srba}
\end{figure}

To classify our stars and learn about the origin(s) of their neutron-capture elements, we plot [Sr/Ba] versus [Ba/Fe] in Figure~\ref{fig:srba}. This offers a way to empirically learn about the nucleosynthesis processes that may have produced these elements, as it remains unclear what underlying nucleosynthesis processes cause the observed differences between typical halo stars and stars in UFDs. As has been shown in \cite{Frebel2014_Segue} and \cite{Ji19}, in this type of plot, the UFD stars are removed from the halo stars that form the main branch. Instead, they form their own branch at lower [Sr/Ba] values. This clustering is immediately evident in Figure~\ref{fig:srba}.

We find that three of our stars, HE~2303$-$5756, HE~2155$-$2043, and HE~0104$-$5300, are located along the main halo branch. Meanwhile, the three other stars, HE~2319$-$5228, HE~1310$-$0536, and HE~2340$-$6036, are squarely within the region occupied by the UFD stars. This telling distinction between the two halves of our sample stars, as shown in Figure~\ref{fig:srba}, suggests that three of our stars are similar to stars in UFDs (hence the square symbols) while the other three are similar to typical halo stars (hence the triangle symbols).

In line with our broader goal of building a set of metal-poor halo stars that possibly originated from small satellite galaxies, we searched the literature for objects similar to our UFD star-like targets (square symbols in Figure~\ref{fig:srba}). Using JINAbase \citep{Abohalima2018_JINAbase}, we specifically looked for halo stars known to have low Sr abundances, specifically $\mbox{[Sr/H]}<-4.5$. We found 61 of them (detailed in Table~\ref{tab:sass_stars}), which we added as orange diamonds to the bottom panel of Figure~\ref{fig:srba} \citep{Li15, Barklem05, Roederer14c, Cayrel04, Hansen15, Lai08, placco14_bd, hollek11, Jacobson15, cohen13, norris07, bonifacio09, ryan96, Col06, Yong13a, keller14, frebel15b, caffau11, cohen08, Frebel07, Mardini_CD-24}. These panels are identical except for the addition of literature stars on the bottom panel. We opted to show these two datasets separately for clarity. Further discussion on the literature stars is presented in Section~\ref{sec:discussion}.

\begin{figure}
\hspace{-0.6cm}
    \includegraphics[width=1\linewidth]{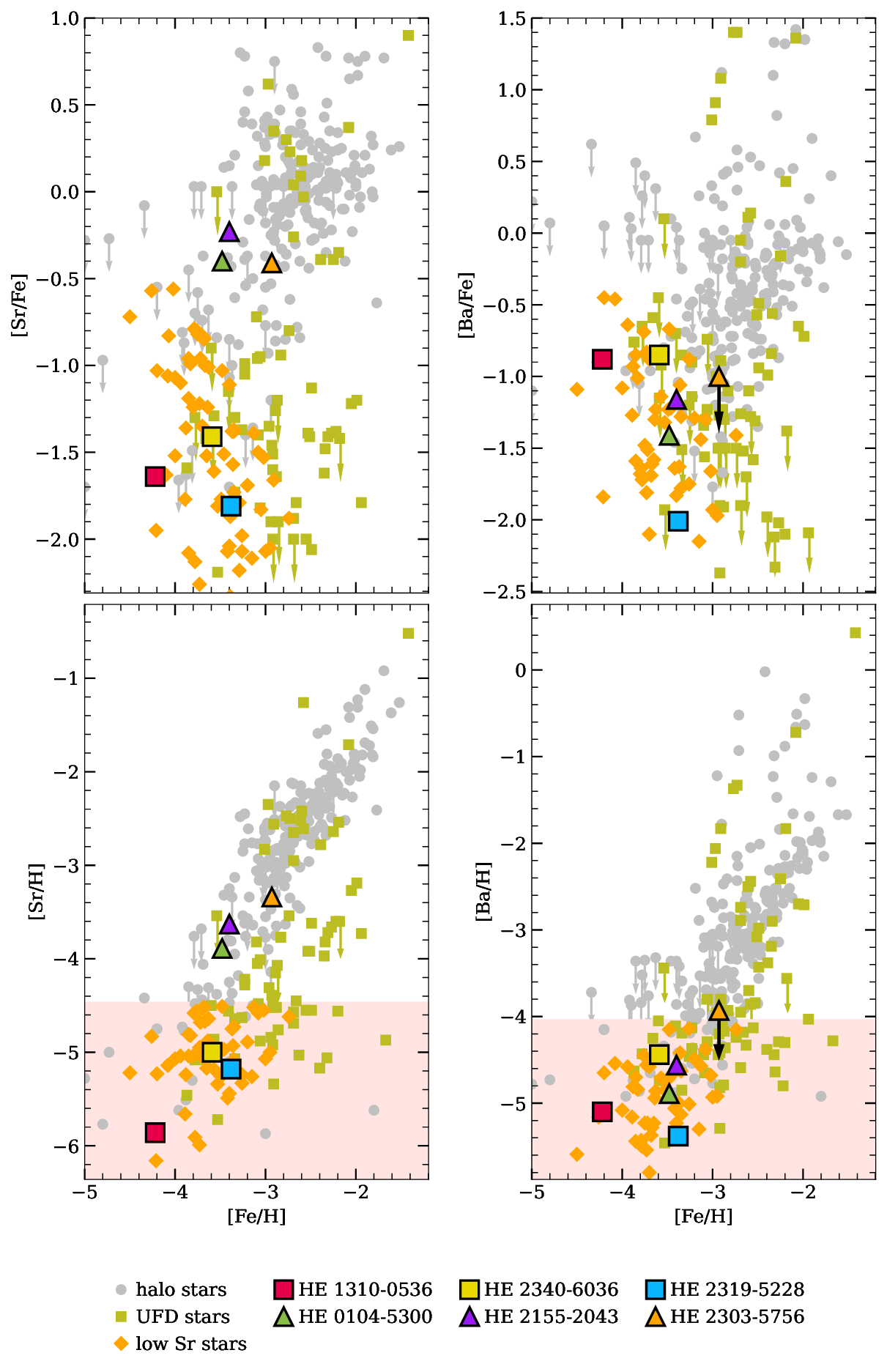} 
        \caption{Sr and Ba abundances for our sample stars in comparison to halo stars (gray points), UFD stars (olive squares), and low Sr stars in the halo (orange diamonds). The program stars denoted by square symbols are those with [Sr/Ba] and [Ba/Fe] ratios similar to those of UFD stars. Meanwhile, the program stars denoted by triangles have halo-like [Sr/Ba] and [Ba/Fe] ratios. The pink-shaded regions on the bottom panels guide the eye to the lowest neutron-capture abundances (see Section~\ref{sec:disc_SASS} for discussion).} 
        \label{fig:ncap abundances}
\end{figure}

We then considered the overall abundance trends of Sr and Ba in various stars. We chose to focus on [Sr/H] and [Ba/H] values in our plots as these abundance ratios provide nucleosynthetically decoupled axes (as opposed to [Sr/Fe] and [Ba/Fe]). For completeness, we nevertheless show [Sr/Fe] and [Ba/Fe] values. In Figure~\ref{fig:ncap abundances}, we show our stars' Sr and Ba abundances, in comparison with a general set of metal-poor halo stars (gray points; same set as in Figure~\ref{fig:light abundances}), with metal-poor stars in various UFDs (olive squares; same set as in Figure~\ref{fig:light abundances}), and with the halo stars known to have low Sr abundances (orange diamonds; same set as in Figure~\ref{fig:srba}). 

In Figure~\ref{fig:ncap abundances}, we keep the same shape grouping as derived in Figure~\ref{fig:srba} and explained in Section~\ref{sec:heavy}. For [Sr/H], the square-denoted stars are clustered away from the triangle-denoted stars, lying at the bottom end of the overall trend set by halo stars. They also have slightly lower [Fe/H] values than the triangle-denoted stars. However, for [Ba/H], there is no separation between the two groups. They all cluster towards the lowest Ba abundances as set by the halo stars. As noted briefly in Section~\ref{sec:light iron trends}, the shape grouping also seems to correlate with light element abundances. We discuss this behavior further in Section~\ref{sec:discussion}. 

We note for completeness that we detected no Eu lines in any of our stars but we provide upper limits in Table~\ref{tab:chemresults}.

\subsection{Abundance uncertainties}

From our measurements of individual spectral lines for each element and the corresponding standard deviations $\sigma$, we calculated standard errors for each of our abundance measurements. However, for elements with a small number of lines ($2 < N < 30$), the resulting standard error will be underestimated. We thus applied small $N$ corrections to arrive at our final values. We list our standard errors in Table~\ref{tab:chemresults}. In several cases, however, the standard error was unrealistically small ($<0.05$\,dex). For these cases, we thus adopt a minimum nominal abundance uncertainty of 0.05\,dex, a figure informed by the $S/N$ and other typical uncertainties. For elements with only a single line detection, we estimated an uncertainty between $0.10-0.15$\,dex depending on the quality of the local spectrum and the $S/N$. Generally, our uncertainties range between $0.05-0.30$\,dex, reflecting the relatively good data quality.  For elements without any line detections, we estimated their abundances with a 3$\sigma$ upper limit, without including any uncertainty.

To determine the systematic uncertainties in our measurements, we varied each stellar parameter by its uncertainty to assess how the abundance changes in response to changes in stellar parameter values. We varied the effective temperature by $100$\,K, $\log g$ by $0.3$\,dex, and the microturbulence by $0.3$\,dex. We then recorded the change in abundance for each element. Our uncertainties are listed in Table~\ref{tab:abundunc} for HE~2155$-$2043 and HE~2303$-$5756, two example stars that illustrate this effect. The total uncertainties were found by adding all of the previously measured uncertainties in quadrature. Generally, the total uncertainties range from 0.10 to 0.30\,dex.

\subsection{Abundance comparison with literature values}

In Table~\ref{tab:comps}, we compare our abundance results with those from previous works that already analyzed our program stars. For HE~2155$-$2043, we compare our measurements with those from \citet{Purandardas} as well as the handful of measurements provided in \citet{Hansen2016}. For HE~0104$-$5300, we compare our results with those from \citet{Jacobson15}. 

The manner in which each paper derived surface gravity varied, thus leading to differing numbers between this study and the others. This explains, for instance, the large discrepancy in [Mg/Fe] between our work and \citet{Purandardas}. There is also a significant difference in [C/Fe], likely because we normalized the carbon G-band region of our respective spectra differently. We tested several different normalization schemes but we were not able to reproduce their reported value of [C/Fe] $\sim 2$. Meanwhile, \citet{Hansen2016} reported [C/Fe] for HE~2155$-$2043 and found a value closer to ours. 

As for the comparison with \citet{Jacobson15}, our measurements for HE 0104$-$5300 agree reasonably well. Overall, for both stars, there is a moderate to good agreement between our results and the literature values at the $0.2- 0.3$\,dex level.

\begin{table}
    \centering
    \caption{Example abundance uncertainties for HE~2155$-$2043 and HE~2303$-$5756}
    \label{tab:abundunc}
    \begin{tabular}{lrrrrr}
    \hline \hline
   Species & Statistical & $\Delta T_{\text{eff}}$ & $\Delta\log(g)$ & $\Delta v_{\text{mic}}$ & Total \\
   & Unc. & $[+100$ K] & $[+0.3$ dex] & $[+0.3$ dex] & Unc. \\
    
    \hline
    \multicolumn{6}{c}{HE~2155$-$2043} \\
    \hline
    C (CH) & 0.10 & $-$0.11 & $-$0.12 & 0.01 & 0.19 \\
    \ion{Na}{i} & 0.21 & 0.09 & $-$0.06 & $-$0.15 & 0.27 \\
    \ion{Mg}{i} & 0.05 & 0.09 & $-$0.07 & $-$0.08 & 0.15 \\
    \ion{Al}{i} & 0.10 & 0.25 & $-$0.07 & $-$0.11 & 0.30 \\
    \ion{Si}{i} & 0.20 & 0.06 & 0.06 & $-$0.13 & 0.25 \\
    \ion{Ca}{i} & 0.05 & 0.07 & $-$0.03 & $-$0.01 & 0.09 \\
    \ion{Sc}{ii}& 0.05 & 0.06 & 0.08 & $-$0.01 & 0.11 \\
    \ion{Ti}{i} & 0.08 & 0.12 & $-$0.02 & $-$0.01 & 0.15 \\
    \ion{Ti}{ii}& 0.05 & 0.07 & 0.10 & $-$0.06 & 0.14 \\
    \ion{Cr}{i} & 0.05 & 0.12 & $-$0.02 & $-$0.03 & 0.13 \\
    \ion{Mn}{i} & 0.15 & 0.16 & 0.02 & $-$0.03 & 0.22 \\
    \ion{Fe}{i} & 0.05 & 0.12 & $-$0.03 & $-$0.08 & 0.16 \\
    \ion{Fe}{ii}& 0.05 & 0.02 & 0.10 & $-$0.01 & 0.11 \\
    \ion{Co}{i}& 0.05 & 0.09 & $-$0.02 & $-$0.07 & 0.13 \\
    \ion{Ni}{i} & 0.05 & 0.13 & $-$0.05 & $-$0.15 & 0.21 \\
    \ion{Zn}{i} & 0.21 & 0.07 & 0.05 & 0.01 & 0.23 \\
    \ion{Sr}{ii}& 0.06 & 0.10 & 0.14 & $-$0.27 & 0.32\\
    \ion{Ba}{ii}& 0.08 & 0.09 & 0.08 & 0.00 & 0.14 \\
    \hline
    \multicolumn{6}{c}{HE~2303$-$5756} \\
    \hline
    C (CH) & 0.35 & 0.15 & 0.15 & 0.13 & 0.43 \\
    \ion{Na}{i} & 0.09 & 0.07 & $-$0.01 & 0.01 & 0.11 \\
    \ion{Mg}{i} & 0.05 & 0.05 & $-$0.03 & 0.00 & 0.08 \\
    \ion{Al}{i} & 0.05 & 0.02 & 0.00 & 0.01 & 0.05 \\
    \ion{Si}{i} & 0.15 & 0.10 & 0.00 & 0.00 & 0.18 \\
    \ion{Ca}{i} & 0.05 & 0.06 & $-$0.01 & 0.01 & 0.08 \\
    \ion{Sc}{ii}& 0.05 & 0.05 & 0.09 & 0.02 & 0.12 \\
    \ion{Ti}{i} & 0.05 & 0.08 & $-$0.01 & 0.02 & 0.10 \\
    \ion{Ti}{ii}& 0.05 & 0.05 & 0.08 & 0.00 & 0.11 \\
    \ion{Cr}{i} & 0.05 & 0.10 & $-$0.01 & 0.02 & 0.11 \\
    \ion{Mn}{i} & 0.15 & 0.03 & $-$0.07 & $-$0.04 & 0.17 \\
    \ion{Fe}{i} & 0.05 & 0.08 & $-$0.01 & 0.01 & 0.10 \\
    \ion{Fe}{ii}& 0.05 & 0.02 & 0.11 & 0.01 & 0.12 \\
    \ion{Co}{i} & 0.05 & 0.09 & $-$0.04 & 0.00 & 0.11 \\
    \ion{Ni}{i}& 0.05 & 0.13 & $-$0.05 & 0.02 & 0.15 \\
    \ion{Sr}{ii}& 0.06 & $-$0.03 & 0.04 & $-$0.05 & 0.09 \\\hline
    \end{tabular}
\end{table}

\begin{table}\centering
\caption{Comparisons with results from other studies}
    \label{tab:comps}
    \begin{tabular}{lrrrrr}
     \multicolumn{6}{c}{HE 2155$-$2043} \\
    \cline{1-6}
    & \multicolumn{2}{c}{Purandardas} &  \multicolumn{1}{c}{Hansen}  & \multicolumn{2}{c}{This Study} \\
    & \multicolumn{2}{c}{et al. 2021} &  \multicolumn{1}{c}{et al. 2016}  & \multicolumn{2}{c}{} 
    \\
     \cline{2-3} \cline{4-4}  \cline{5-6}
    Species & $\log\epsilon$(X) & [X/Fe] & [X/Fe] & $\log\epsilon$(X) & [X/Fe] \\ 
    \hline
    \hline
    C (CH) & 7.05 & 2.05 & 0.7 &  6.00  & 0.97\\ 
    \ion{Na}{i} & 3.79 & 0.98 & ... & 3.37 & 0.54 \\ 
    \ion{Mg}{i}  & 5.74 & 1.57 &  ... & 4.72 & 0.53 \\ 
    \ion{Al}{i} &...  & ... &...  & 2.29 & $-$0.76 \\ 
    \ion{Si}{i} & ... & ... & ... & 4.06 & $-$0.04 \\ 
    \ion{Ca}{i} & 3.35 & 0.44 &...  &  3.32 & 0.38 \\ 
    \ion{Sc}{ii} & $-$0.36 & $-$0.02 & ... &  $-$0.22 & $-$0.03 \\ 
    \ion{Ti}{i} & ... & ... & ... &  1.93  & 0.38 \\ 
    \ion{Ti}{ii} & 1.67 & 0.21 & ... & 1.70 & 0.15 \\
    \ion{Cr}{i} & 2.11 & $-$0.10 & ... &  1.93 & $-$0.30\\
    \ion{Mn}{i} & 2.04 & 0.04 &  ... & 1.29 & $-$0.73 \\
    \ion{Fe}{i} & 4.07 & 0.00 & 0.00 &  4.10 & 0.00 \\
    \ion{Fe}{ii} & 4.01 & ... & ... &  4.18 & 0.08\\
    \ion{Co}{i} & 2.46 & 0.90 & ... & 2.05 & 0.47 \\
    \ion{Ni}{i} & 3.09 & 0.30 & ... & 2.92  & 0.10 \\
    \ion{Zn}{i} &...  &...  &...  &  1.77 & 0.61\\
    \ion{Sr}{ii} & $-$0.66 & $-$0.04 & 0.2 & $-$0.76 & $-$0.23 \\
    \ion{Ba}{ii} & $-$2.95 & $-$1.64 &  ... & $-$2.38 & $-$1.16\\
    \ion{Eu}{ii} & ... & ... & ... &  $<-$2.50 & $<$0.38\\
    \hline
\end{tabular}

\vspace*{0.5 cm}
    
    \begin{tabular}{lrrrr}
     \multicolumn{5}{c}{HE 0104$-$5300} \\
    \cline{1-5}
    & \multicolumn{2}{c}{Jacobson} & \multicolumn{2}{c}{This Study} \\
    & \multicolumn{2}{c}{et al. 2015} & \multicolumn{2}{c}{} 
    \\
     \cline{2-3} \cline{4-4}  \cline{5-5}
    Species & $\log\epsilon$(X) & [X/Fe] & $\log\epsilon$(X) & [X/Fe] \\ 
    \hline
    \hline
    C (CH) & 4.77 & 0.13   & 4.77 & $-$0.17 \\ 
    \ion{Na}{i} & 2.74 & 0.29 & 2.94 & 0.18 \\ 
    \ion{Mg}{i} & 4.37 & 0.57  & 4.50 & 0.38 \\ 
    \ion{Al}{i} & 2.06 & $-$0.60  & 2.37 & $-$0.60 \\ 
    \ion{Si}{i} & 4.65 & 0.93   &4.67 & 0.64 \\ 
    \ion{Ca}{i} & 3.01 & 0.46   & 3.22 & 0.36 \\ 
    \ion{Sc}{ii} & $-$0.83 & $-$0.19   & $-$0.29 & 0.05 \\ 
    \ion{Ti}{i} & 1.37 & 0.21   & 1.74 & 0.27 \\ 
    \ion{Ti}{ii} & 1.41 & 0.25   & 1.91 & 0.45 \\
    \ion{Cr}{i} & 1.39 & $-$0.46   & 1.89 & $-$0.26 \\
    \ion{Mn}{i} & 1.12 & $-$0.52   & 1.07 & $-$0.88 \\
    \ion{Fe}{i} & 3.71 & 0.00 & 4.02 & 0.00 \\
    \ion{Fe}{ii} & 3.72 & 0.01  & 4.27 & 0.25 \\
    \ion{Co}{i} & 1.34 & 0.14   & 1.82 & 0.31 \\
    \ion{Ni}{i} & 2.30 & $-$0.12   & 2.92 & 0.18 \\
    \ion{Zn}{i} & ... & ... &  1.64 & 0.56 \\
    \ion{Sr}{ii} & $-$1.71 & $-$0.79   & $-$1.02 & $-$0.40 \\
    \ion{Ba}{ii} & $-$3.25 & $-$1.64   & $-$2.71 & $-$1.41 \\
    \ion{Eu}{ii} & ... & ... &   $<-$2.62 & $<$0.34 \\
    \hline
\end{tabular}
\end{table}


\section{Kinematic Analysis} \label{sec:kinematicanalysis}

In addition to chemical abundances, we also use stellar kinematics to reconstruct each star's orbital history and learn more about its origin scenario. We employ the same procedure detailed in \citet{Mardini2022_J1808}. The authors used \textit{Gaia} astrometric data in combination with a time-varying Galactic potential modeled in The \textsc{orient}\footnote{Available at \url{https://github.com/Mohammad-Mardini/The-ORIENT}}. The potential was constructed from snapshots of a large-scale cosmological simulation of a Milky Way-like halo (see \citealt{Mardini2020} for more details).

For this work, we determined distances to our sample stars using data from \textit{Gaia} DR3 \citep{GaiaDR3}. Despite \textit{Gaia}'s unprecedented precision, a considerable number of its parallax measurements have large uncertainties such that distances derived by simply inverting the parallax ($d=\frac{1}{p}$) can be unreliable. To correct this, we adopt the method described in \cite{Mardini2022_Atari} where we first employ a zero-point correction using a pseudocolor where available. We then probabilistically derive a final distance estimate using a space density prior, also in combination with the zero-point correction. 

However, as \cite{Luri2018_parallax} recommended, distances should ideally be derived using a full Bayesian approach. Following this suggestion, we thus infer a final distance measurement using the more rigorous probabilistic method described by \cite{Bailer-Jones2018_SDP}. As such, we adopt an exponentially decreasing space density prior (SDP; as opposed to an isotropic prior) to calculate our final distance estimate. We then run a Monte Carlo simulation generating 10,000 realizations for each star and assuming a normal distribution centered around the corrected parallax and its standard deviation. The mean of the resulting posterior distribution is the final distance for each star. Our final distances are presented in Table~\ref{tab:kinematicparams}. 

We then calculate the positions and velocities of our sample stars using \textit{Gaia} coordinates, proper motion, and radial velocity. We adopt a position of the Sun of $R_{\odot}=8.178\pm0.013$ kpc away from the Milky Way's center \citep{Gravity2019_MWcenter} at a vertical distance $z_{\odot}=20.8\pm0.3$\,pc above the Galactic plane and solar velocities as given by \citet{BennettBovy2019_Udot} and \citet{Schonrich2010}. We adopt $V_{\text{LSR}}=220$ \unit{ km.s^{-1}} as measured by \citet{KerrLyndenBell1986_VLSR}. Table~\ref{tab:kinematicparams} summarizes the kinematic parameters of our sample stars. 

The strong retrograde $V$ motions of all of our stars (i.e., $V < 0$), especially those of HE~2303$-$5756 and HE~2155$-$2043 with $V \sim -450$ \unit{ km.s^{-1}}, suggest that our sample is indeed composed of halo stars, a finding which aligns with their extremely metal-poor nature. These results suggest a possible early accretion origin of our stars. We also tested using $V_{\text{LSR}}=232$ \unit{km.s^{-1}} \citep{McMillan17_MWPot} and $V_{\text{LSR}}=238$ \unit{km.s^{-1}} \citep{Schonrich12} to assess the impact of different $V_{\text{LSR}}$ values on our conclusions regarding the origins of our stars. Using these new values changes $V$ by only 10\,\unit{km.s^{-1}}. Thus, changing $V_{\text{LSR}}$ does not affect our conclusion regarding the retrograde motion of our stars.

\begin{figure*}
\centering
    \includegraphics[width=1\linewidth]{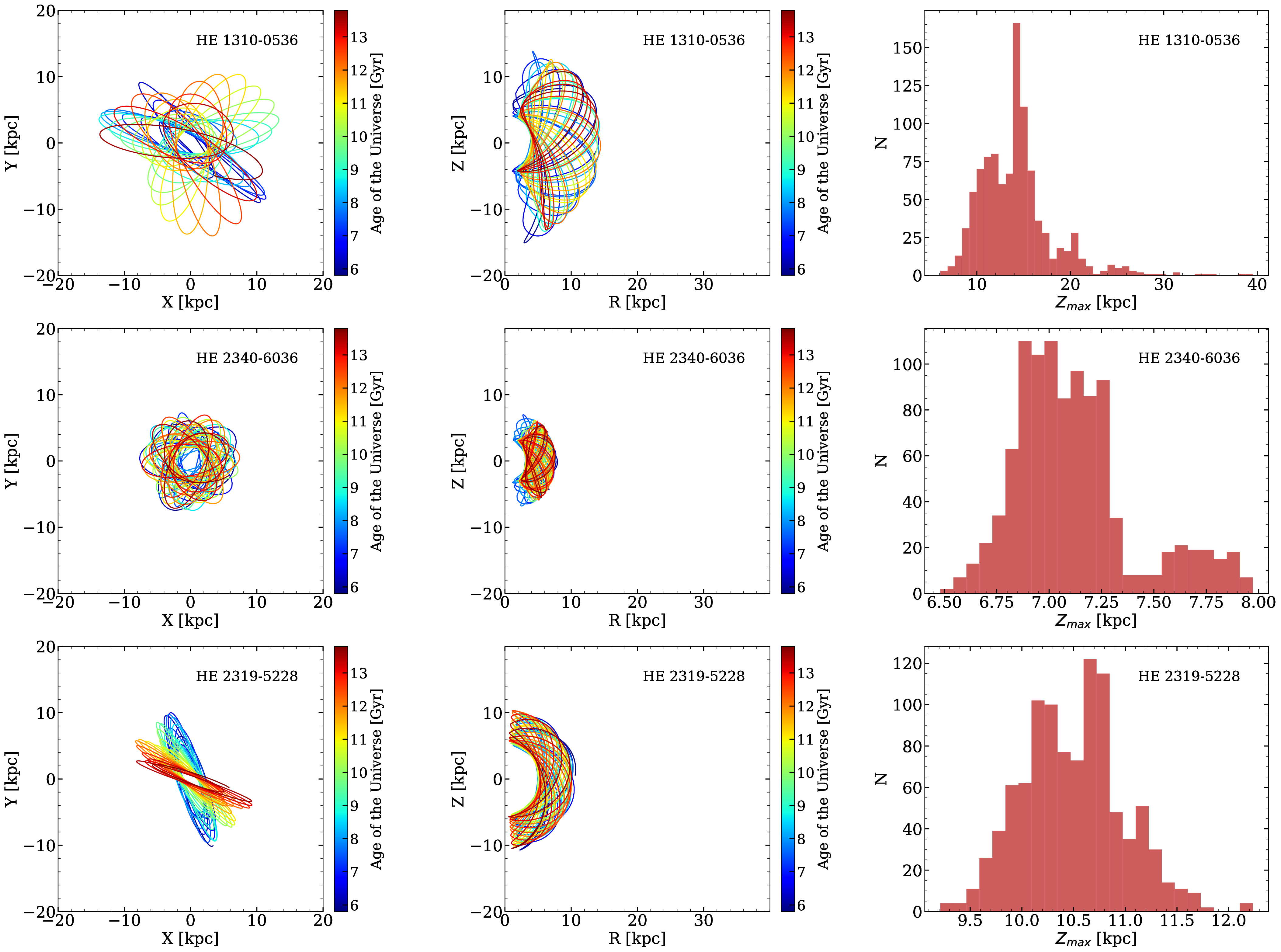}
    \caption{Orbital evolution over the last 8 billion years of our three UFD star-like targets. For every star, we show one realization of its orbital history, back-integrated using the time-dependent \textsc{orient} potential \#483868. Left: View of the Galactic plane from above in X-Y coordinates (in kpc). Center: Side view showing the Z height of the orbit as a function of distance from the Galactic Center. Right: Histogram showing the distribution of $Z_{\text{max}}$ (maximum distance above or below the Galactic plane) over 1000 realizations of the star's orbital evolution.}
    \label{fig:orbits}
\end{figure*}

Given all these inputs, we derive the Galactocentric coordinates $(X, Y, Z)$ with a right-handed frame and rectangular Galactic velocities $(U, V, W)$. With these kinematic parameters, we then back-integrate each star's orbit with a time-varying Galactic potential, namely \textsc{orient} potential \#483868 \citep{Mardini2022_J1808}. To investigate their orbits statistically, we generate 1,000 realizations of each star's orbital evolution. In Figure~\ref{fig:orbits}, we show one such realization for our UFD star-like targets, namely HE~1310$-$0536, HE~2340$-$6036, and HE~2319$-$5228 (square symbols in relevant figures).

Of particular note is each star's $Z_{\text{max}}$, its maximum distance above or below the Galactic plane. On the right column of Figure~\ref{fig:orbits}, we show the distribution of $Z_{\text{max}}$ values over 1,000 realizations. In Table~\ref{tab:kinematicparams}, we list our program stars' median $Z_{\text{max}}$ values. All of our program stars have significant $Z_{\text{max}}$, generally ranging between $7-48$\,kpc, once again confirming that these are halo stars. Many of them also have rather elliptical orbits about the Galactic Center.

For HE~1310$-$0536, we were unable to derive a reliable distance due to large uncertainties in the \textit{Gaia} data ($d = 11.9\pm2.3$ kpc). To err on the side of caution, we thus carried out our kinematic analysis (described above) of  HE~1310$-$0536 three times, with three possible distances taking into account its uncertainty (9, 12, and 14\,kpc). This was to test the effects of the distance uncertainty on our orbital history results. We find that in all three cases, the star appears to be part of the halo, rather than the disk.

\begin{table*}
    \centering
    \caption{Kinematic parameters of our sample stars}   \label{tab:kinematicparams}
    \begin{tabular}{Xlrrrrrrrr}
    \hline \hline
   Star & $X$ & $Y$ & $Z$ & $U$ & $V$ & $W$ & Distance & $Z_{\text{max}}$\\
   & [kpc] & [kpc] & [kpc] & [\unit{km.s^{-1}}] & [\unit{km.s^{-1}}] & [\unit{km.s^{-1}}] & [kpc] & [kpc]\\
    \hline
    HE 0104$-$5300 & $-6.73$ & $-2.81$ & $-6.54$ & $-177.98$ & $-303.15$ & $-117.25$ & $7.4\pm0.6$ & 31 \\
    HE 1310$-$0536 & $-3.88$ & $-4.60$ & $9.51$ & $-137.98$ & $-263.95$ & $64.35$ & $11.9\pm2.3$ & 14 \\ 
    HE 2155$-$2043 & $-4.11$ & $2.64$ & $-5.81$ & $-216.27$ & $-467.79$ & $-245.61$ & $6.4\pm1.2$ & 48 \\
    HE 2303$-$5756 & $-7.65$ & $-0.34$ & $-0.86$ & $-213.62$ & $-430.29$ & $155.35$ & $1.1\pm0.0$ & 12 \\
    HE 2319$-$5228 & $-5.97$ & $-1.25$ & $-4.37$ & $57.12$ & $-190.84$ & $-245.15$ & $5.3\pm0.4$ & 11\\
    HE 2340$-$6036 & $-6.10$ & $-1.92$ & $-4.02$ & $124.93$ & $-281.53$ & $-60.58$ &$5.6\pm0.4$ & 7\\
    \hline
    \end{tabular}
\end{table*}

\begin{figure*}
\centering
    \includegraphics[width=1\linewidth]{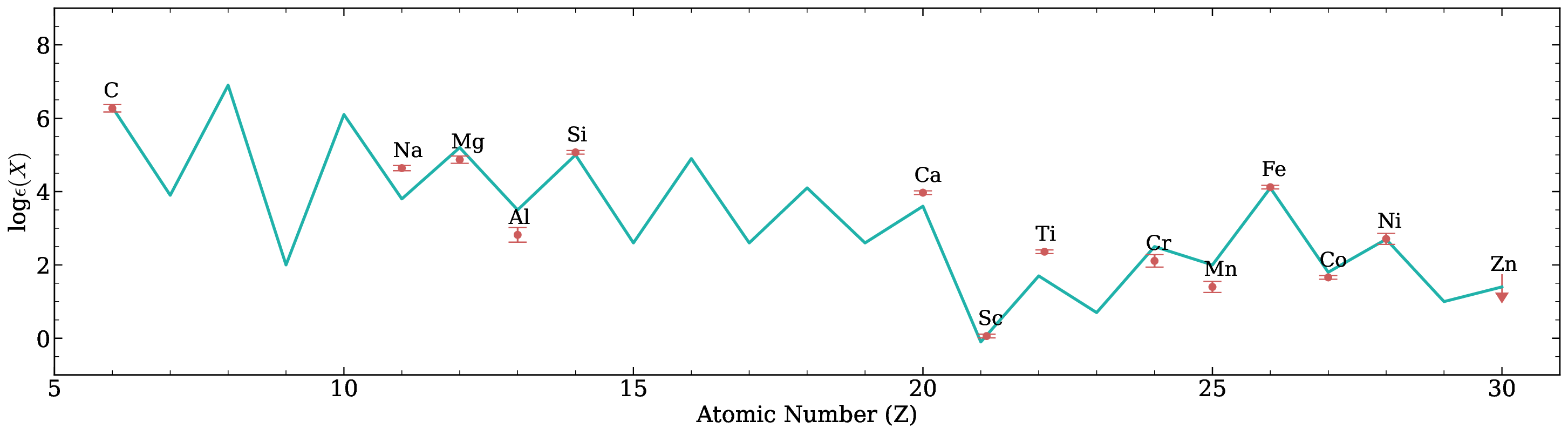}
    \caption{Abundances of HE$~$2319$-$5228 as a function of atomic number. Overlaid is a Starfit supernova nucleosynthesis model (teal line) that best matches the data (red points). Sc is not considered in the fit given uncertainties in the Sc production, as recommended by Starfit. Overall, the fit is moderate given that Cr and Mn are significantly overproduced by the model. It corresponds to a progenitor with $\sim$27 solar masses and an explosion energy of 0.9 Bethe. Applying exemplary NLTE corrections to multiple elements did not significantly change the fitting outcome.}
    \label{fig:2319_starfit} 
\end{figure*}


\section{On the origins of halo stars with low neutron-capture element abundances} \label{sec:discussion}

\subsection{Small Accreted Stellar System (SASS) stars} \label{sec:disc_SASS}

In Figure~\ref{fig:light abundances}, we showed that our six sample stars — all metal-poor and deficient in neutron-capture elements — generally follow the light element abundance trends set by typical halo stars. In Section~\ref{sec:kinematicanalysis}, we also showed that all our stars have halo kinematics with a strong propensity for significantly retrograde orbits. Yet, when considering their [Sr/Ba] vs. [Ba/Fe] behavior (as shown in Figure~\ref{fig:srba}), the sample splits into two distinct groups. Three stars fall in the region typically occupied by UFD stars, while the other three stars overlap with the main halo star branch. This division persists noticeably in the [Sr/H] vs. [Fe/H] panel in Figure~\ref{fig:ncap abundances}. To some degree, this division is also seen in some of the light elements, particularly in Ca, Co, and Ni. 

Generally, stars in UFDs have a strong tendency to cover regions of the lowest [Sr/H] and [Ba/H] values, i.e., [Sr/H] $<-4.0$ and [Ba/H] $<-4.0$ as shown in \cite{Frebel2014_Segue}. To arrive at more stringent results, we restrict the following discussion to an even lower neutron-capture abundance cutoff at [Sr/H] $<-4.5$ and [Ba/H] $<-4.0$. These low Sr and Ba regions are colored pink in Figure~\ref{fig:ncap abundances} to guide the eye. 

Indeed, half of our sample stars (denoted with squares for their UFD-like [Sr/Ba] vs. [Ba/Fe] behavior) fall in the pink low-Sr region in the [Sr/H] versus [Fe/H] panel of Figure~\ref{fig:ncap abundances}. Meanwhile, the other three stars (denoted with triangles) do not. For the [Ba/H] versus [Fe/H] panel, however, all sample stars more or less cluster in the pink low-Ba region typically covered by UFD stars. All these stars are extremely metal-poor.

Altogether, the three characteristics of (1) low [Fe/H] values paired with a corresponding [Sr/Ba] value (as in Figure~\ref{fig:srba}), (2) low [Sr/H] values, and (3) retrograde halo orbits suggest that three of our stars have origins extremely similar to those of typical UFD stars. This similarity implies that these stars (HE 2319$-$5228, HE 2340$-$6036, HE 1310$-$0536) may have originated from small dwarf galaxies long accreted by the Milky Way. On the other hand, the other three stars (HE 0104$-$5300, HE 2155$-$2043, HE 2303$-$5756) may have different origin scenarios. Perhaps they came from larger dwarf galaxies (analogous to the classical dwarf spheroidal galaxies) which underwent some amount of chemical evolution before eventually being accreted by the Galaxy. Their Sr and Ba behavior resembles that of several extant satellite systems (e.g., \citealt{Venn04}). 

Informed by the above arguments and by observations of stars in UFDs, we can now assume that stars with the lowest Fe and Sr values originated in some of the smallest, earliest systems that formed in the universe. We will refer to these stars (the low-Sr stars represented as squares in the various figures) as stars from Small Accreted Stellar Systems (SASS). Now we will go one step further and assume that if a star in the Galactic halo has an extraordinarily low Sr abundance, then it must also be a SASS star. As will be discussed later, we indeed find many such SASS stars in the literature. Together, these stars open up new ways of studying early star formation environments, the early assembly history of the Milky Way, and the population of the earliest galaxies. 

In the following subsections (\ref{disc_light}, \ref{disc_carbon}, \ref{disc_snyields}, \ref{disc_ncap}), we discuss the environments of early dwarf galaxies and establish that SASS stars indeed originated from such environments. In later subsections (\ref{disc_SASSlit}, \ref{disc_sass_earlyuniverse}), we discuss SASS stars in the literature and the broader potential of SASS stars as probes of the early universe.

\subsection{Light element abundances} \label{disc_light}

As noted in Subsection~\ref{sec:light iron trends}, SASS stars tend to show larger deviations from the well-established halo trends in their light element abundances — a behavior found repeatedly among the most iron-poor stars. These iron-poor stars are thought to have formed from gas that was enriched by individual and/or unusual supernova progenitors and/or gas that underwent inhomogeneous mixing \citep{Christlieb04,frebel05,placco15, frebel15}. This behavior offers principal support for our claim that SASS stars formed in the earliest systems; they experienced enrichment events that were individual and more stochastic.

\subsection{Carbon abundances} \label{disc_carbon}

Another feature of the most iron-poor stars is strong carbon enhancement. Two of our three SASS stars are strongly carbon-enhanced, with HE 2319$-$5228 having [C/Fe] $=1.2$ and HE 1310$-$0536 having [C/Fe] $=2.7$. Once again, this lends further support to our claim that SASS stars came from the earliest systems. However, our third SASS star, HE~2340$-$6036, has a low carbon abundance with [C/Fe] $=-0.55$. The origin of such a low value generally remains unclear, but it could be due to extreme inhomogeneous mixing. 

Interestingly, stars in UFDs tend to have only mild carbon enhancement or even subsolar levels of carbon \citep{frebel16, Norris10b}. Only one of our SASS stars, HE~2340$-$6036, has UFD-like carbon abundance levels. Meanwhile, our other two SASS stars are grossly different in that they are much more carbon-enhanced. We note, however, that several individual UFD stars are known to have such extremely high carbon abundances \citep{LAI11}. More data might help shed light on the overall C behavior in UFDs. In the meantime, we continue to assume that SASS stars are indeed not unlike the surviving UFD stars.

\subsection{Supernova nucleosynthesis yields} \label{disc_snyields}

We further test the idea of early pre-enrichment by considering supernova nucleosynthesis yields and how they match with the now-observed abundances in our SASS stars. We used Starfit\footnote{Available at \url{https://starfit.org}} \citep{Heger2010}, a fitting tool that matches the observed light element ($Z<30$) abundance signatures with models of Pop\,III supernova nucleosynthesis yields. This provides clues to the properties of the (possibly Pop\,III) progenitors that operated in the birth environments of our SASS stars. 

Studies of the ultra-faint dwarf galaxies Hercules and Leo IV suggest that they may have been enriched by the supernovae of putative Population III stars \citep{Koch08, Simon2010_LeoIV}. We nominally find Pop\,III progenitor masses of 10 to 23\,M$_\odot$ and accompanying explosion energies of $\sim$10$^{50}$ to 10$^{51}$\,ergs. An example fit is shown in Figure~\ref{fig:2319_starfit}. As can be seen, the fit for HE 2319$-$5228 is overall of moderate quality. The same is true for the fits of other stars. These rather low explosion energies are in line with our claim that SASS stars formed in early systems. Otherwise, significantly larger energies might have blown the systems apart and prevented further star formation.

\subsection{Neutron-capture element abundances} \label{disc_ncap}

A remaining question pertains to the origins of low neutron-capture element abundances — a property shared by both UFD and SASS stars. Generally, low neutron-capture element abundances imply either (1) a limited production of said elements, or (2) a limited number of enrichment events in combination with poor mixing and distribution of metals. Possible options for events and processes include the limited $r$-process operating in supernovae \citep{Wanajo06, Arcones11, Frebel2018_ARAA} and other heavy element production processes associated with supernovae such as spinstars
\citep{Maeder15, Hansen2016_CEMPno}, all in concert with the commensurate carbon enhancement events. Crucially, we do not consider neutron star mergers as they would strongly overproduce neutron-capture elements, likely generating abundances higher than those observed in our SASS stars. Not only that, we found upper limits of [Eu/H] $<-$3 for five out of our six program stars. With such low Eu levels, we can reasonably exclude a rare, prolific $r$-process event such as a neutron star merger. Our low Eu levels are more in line with the Eu abundances found in typical UFDs. 

Nevertheless, it is interesting to consider whether any SASS stars have a [Sr/Ba] ratio similar to that produced by either the $s$- or the $r$-process. Typical $r$-process stars have [Sr/Ba] $\sim -0.4$ to $-0.7$ \citep{frebelji23, ji16c} while $s$-process stars have [Sr/Ba] $<-1$. None of our stars has [Sr/Ba] $<-1$. However, one SASS star has [Sr/Ba] =$-0.87$ (HE~1310$-$0536) and another one has [Sr/Ba] =$-0.56$ (HE~2340$-$6036). These stars have extremely low levels of [Sr/H] ($-5.86$ for HE~1310$-$0536; $-5.00$ for HE~2340$-$6036) and [Ba/H] ($-5.10$ for HE~1310$-$0536; $-4.44$ for HE~2340$-$6036). This suggests the potential existence of early $r$-process nucleosynthesis, or at least, a dilution process that led to such low overall levels. Perhaps this nucleosynthetic event was associated with an early supernova such as a magnetorotational hypernova \citep{yong21}. More observations of similar stars could help shed light on this question, which we address further below. For completeness, we note that the third SASS star, HE~2319$-$5228, has [Sr/Ba] = $0.2$. Its neutron-capture elements likely have a different origin.

\subsection{SASS stars in the literature} \label{disc_SASSlit}
\begin{figure*}
        \includegraphics[width=\linewidth]{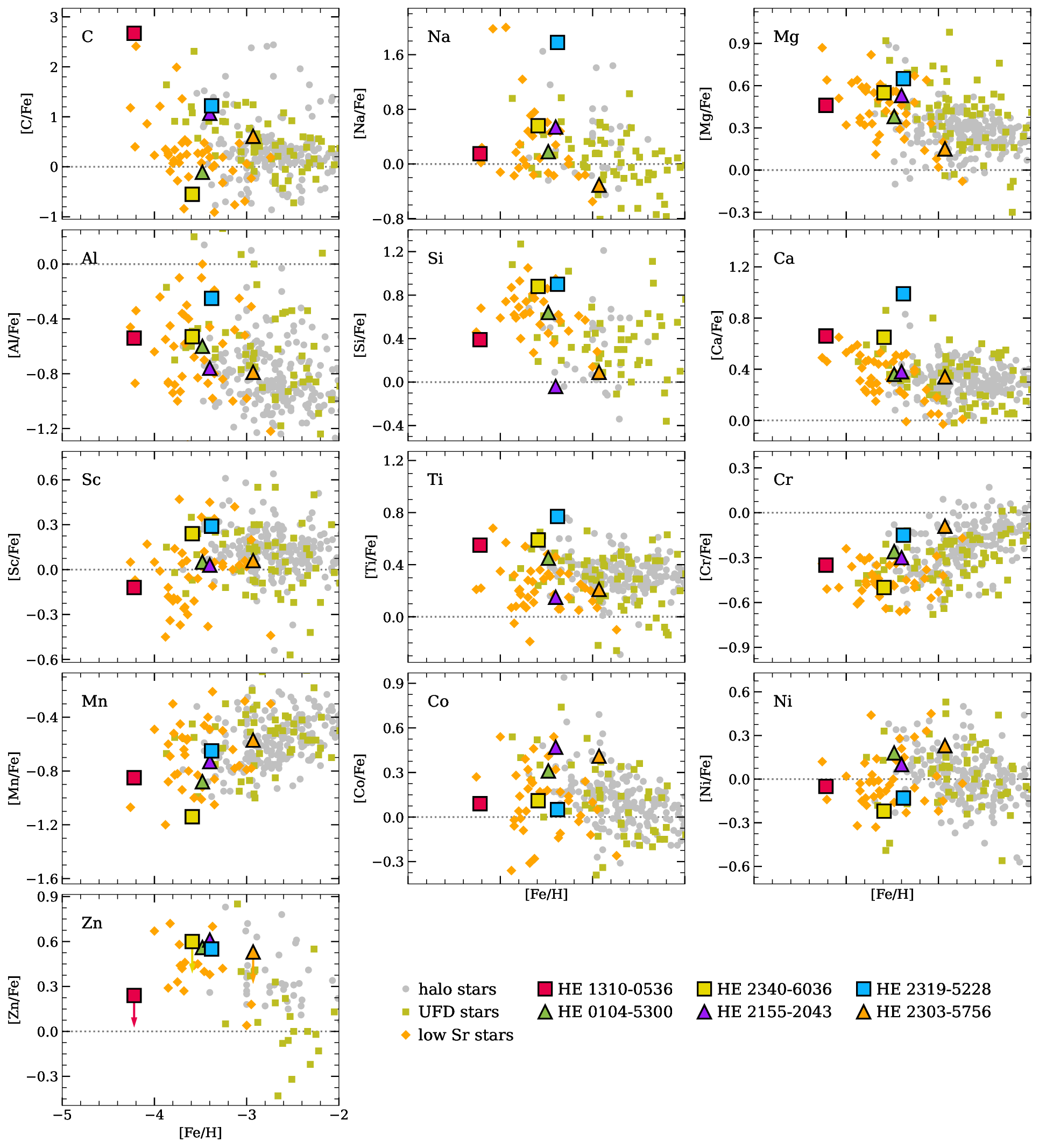} 
        \vspace{-0.5cm}
        \caption{\label{fig:light abundances_lit} Same as Figure~\ref{fig:light abundances} but with halo stars having low [Sr/H] $<-4.5$ added as orange diamonds. While the metallicity is lower for those stars, the scatter is larger than that for the other halo stars. See Section~\ref{disc_SASSlit} for discussion.}         
\end{figure*}

As we established in the previous subsections, our three SASS stars likely originated in the earliest systems in the universe. A large set of such stars would therefore allow us to study in detail the early environments of dwarf galaxies. Motivated by this, we sought to expand our collection of 3 SASS stars by searching the literature for similar stars. To this end, we queried JINAbase for stars with low [Fe/H] and [Sr/H] $<-4.5$. We found 61 such stars which we show as orange diamonds in Figures~\ref{fig:srba}, \ref{fig:ncap abundances}, and \ref{fig:light abundances_lit}. These literature SASS stars are listed in Table~\ref{tab:sass_stars} together with selected chemical abundances, JINAbase identification numbers, and radial velocities. 

In Figure~\ref{fig:ncap abundances}, our literature SASS stars lie within the low-Sr region of the [Sr/H] versus [Fe/H] plot. This is expected since we selected these stars precisely for their having [Sr/H] $<-4.5$. Though these stars were only selected for their low [Sr/H], it is interesting to note that these same stars strictly have low barium abundances; they all have [Ba/H] $<-4.0$. This makes SASS stars generally straightforward to identify using neutron-capture element abundances. However, while a low [Ba/H] value is a necessary characteristic for a SASS star, this criterion by itself is not sufficient for the selection of SASS stars. In addition to being poor in Sr (and by extension, Ba), SASS stars must also have low metallicities, i.e., [Fe/H] $\sim$ $-$5.0 to $-$3.0. This low-metallicity criterion selects for stars that formed in the early universe.

Now we consider the reverse logical path. Looking at their [Sr/Ba] ratio in Figure~\ref{fig:srba}, all these literature SASS stars fall into the UFD region (just as our three program SASS stars). In Figure~\ref{fig:light abundances_lit}, we show the light element abundances of these stars (with abundances as collected from JINAbase). Like our three program SASS stars, the literature SASS stars have a tendency to be more scattered in their light element abundances, compared to the typical halo trend set by stars with higher Sr values. This behavior was foreshadowed by our program SASS stars and is more apparent now with the literature SASS stars. We note that as more stars become available, a more detailed analysis of the overall behavior would deliver more accurate results. 

Finally, we carried out the same kinematic analysis detailed in Section~\ref{sec:kinematicanalysis} on the literature SASS stars to assess their orbital histories. Like our program SASS stars, we find a strong tendency for these stars to be on highly retrograde orbits, with essentially no prograde motion present.

\subsection{Studying the early universe with SASS stars} \label{disc_sass_earlyuniverse}

We thus conclude that halo stars with [Sr/H] $<-4.5$ can be considered SASS stars and that they likely originated from early small systems that were accreted by the Milky Way at the earliest times. After all, the early Milky Way assembled from galactic building block-type systems much like UFDs. Thus it should not be surprising that at least the low-mass stars are still present in our Galaxy and that we can observe them. Accordingly, we suggest using [Sr/H] $<-$4.5 as an important selection tool in identifying the truly oldest stars, namely the SASS stars, in our Galaxy. Furthermore, these stars will be characterized by a suitable UFD-like [Sr/Ba] ratio (together with an appropriate [Ba/Fe] value), as well as a highly retrograde orbit. 

From the chemical abundances of SASS stars, it appears that the earliest systems accreted by the Milky Way were enriched by a very small number of supernovae whose (perhaps unique) yields were somewhat inhomogeneously mixed. This inhomogeneity produced variations in light element abundances, including carbon abundances, from system to system. Neutron-capture elements were sparsely produced and/or diluted likely in multiple ways, with $r$-process nucleosynthesis (at least up to Ba) playing a role. This suggests that most, if not all, elements were produced in small amounts from the earliest times onward. High-resolution cosmological simulations of early (baryonic) structure formation may be able to more fully model the population of the earliest galaxies and building blocks based on these observational constraints. 

Tracing the oldest surviving stars within large galaxies, such as the Milky Way, would be another avenue towards confirming our observational findings that there appears to be a significant population of the very oldest stars present in the halo. Adding the Sr abundance as a criterion in addition to the commonly used Fe abundance may finally offer a path towards identifying not just the most metal-poor but indeed the oldest stars in the universe.

We note that SASS stars may not be the only unique survivors of this very early time. For instance, stars with slightly higher Sr values may well be part of this extremely ancient population. However, we apply a more stringent cut here in this work to ensure the selection of a relatively clean and easily identifiable representative set of this unique population of the very oldest stars in the universe. 

Going forward, SASS stars offer a new path to identifying ancient stars that are relatively close by. These nearby stars, which are easier to observe than the faint and distant dwarf galaxy stars, will allow for more detailed probing of early galactic environments and the galaxy formation era. With many more halo stars available to probe early star formation, we may eventually be able to establish the full history of the systems that built the Milky Way. In addition, we may better understand the conditions of the earliest galaxies which hosted the oldest and most metal-poor stars.

\begin{table*}
    \centering
    \scriptsize
	\caption{\label{tab:sass_stars} SASS stars from the literature}
    \begin{tabular}{lrrrrrrrr}
    \hline \hline
        Name & [Fe/H] & [C/H] & [Mg/H] & [Sr/H] & [Ba/H] & JINAbase ID & RV & Reference \\ \hline
        HE~2248$-$3345 & $-$2.74 & $-$2.55 & $-$2.82 & $-$4.62 & $-$4.15 & 901 &145.4 & BAR05 \\ 
        HE~2249$-$1704 & $-$2.95 & $-$3.18 & $-$2.75 & $-$5.00 & $-$4.92 & 2604 &$-134.1$& COH13 \\ 
           CS29502-042 & $-$3.00 & $-$2.94 & $-$2.98 & $-$5.07 & $-$4.93 & 842 &$-138.05$& CAY04 \\ 
        HE~2304$-$4153 & $-$3.02 & $-$3.71 & $-$2.94 & $-$4.55 & $-$4.68 & 913 &$-221.7$& BAR05 \\ 
        HE~0411$-$5725 & $-$3.08 & $-$2.61 & $-$2.75 & $-$4.58 & $-$4.38 & 2557 &286.1& COH13 \\ 
          HE~0316+0214 & $-$3.13 & $-$3.89 & $-$2.49 & $-$4.52 & $-$4.57 & 268 &$-161.0$& BAR05 \\ 
           BS16084-160 & $-$3.15 & $-$3.23 & $-$3.01 & $-$5.26 & $-$5.30 & 638 &$-132.32$& LAI08 \\
        HE~1416$-$1032 & $-$3.20 & $<-$3.51 & $-$3.04 & $-$4.89 & $-$4.49 & 2623 &$5.2$& COH13 \\ 
        CS22943$-$137    & $-$3.22 & N/A & $-$2.59 & $<-$4.62 & $<-$3.70 & 709 &$-20.0$& RYA96 \\ 
        CD~$-$24{\textdegree}17504& $-$3.23  &$-$1.83  &$-$2.84  &$-$4.75   & $-$4.15&3936&136.4 &MAR24 \\
    LAMOST\,J0006+1057 & $-$3.26 & $-$3.58 & $-$2.59 & $-$5.33 & $-$5.01 & 40 &$-281.6$& LI15a \\ 
SMSS\,J031556.09$-$473442.1 & $-$3.26 & $-$3.07 & $-$2.78 & $-$5.24 & $-$4.14 & 2372 &192.3& JAC15 \\ 
        HE~0008$-$3842 & $-$3.35 & $-$4.26 & $-$3.02 & $-$4.73 & $-$5.13 & 47 &128.9& BAR05 \\ 
        HE~0344$-$0243 & $-$3.35 & $-$3.32 & $-$2.95 & $-$5.08 & $-$4.63 & 2647 &$-110.5$& COH13 \\ 
SMSS\,J003327.36$-$491037.9 & $-$3.36 & $<-$3.76 & $-$2.81 & $-$4.93 & $-$4.42 & 2351 &84.7& JAC15 \\ 
           CS22949-048 & $-$3.37 & $-$3.38 & $-$3.15 & $-$4.75 & $-$5.00 & 925 &$-160.8$& ROE14b \\ 
        CS30492-016    & $-$3.40 & $-$2.65 & $-$2.62 & $<-$5.10 & $<-$3.33 & 1453 &46.0& ROE14b \\
           CS30325-094 & $-$3.40 & $-$3.21 & $-$2.94 & $-$5.44 & $-$5.23 & 596 &$-138.0$& CAY04 \\ 
SMSS\,J023147.96$-$575341.7 & $-$3.42 & $-$3.07 & $-$3.13 & $-$5.49 & $-$5.06 & 1923 &219.2& JAC15 \\ 
        HE~0139$-$2826 & $-$3.46 & $-$2.98 & $-$2.90 & $-$4.97 & $-$4.68 & 2022 &$-253$& PLA14a \\ 
        HE~1347$-$1025 & $-$3.48 & $-$3.21 & $-$3.08 & $-$4.51 & $-$4.15 & 2622 &48.6& COH13 \\ 
        HE~1356$-$0622 & $-$3.49 & $-$3.41 & $-$2.93 & $-$5.26 & $-$4.72 & 2631 &93.5& COH13 \\ 
           CS22942-002 & $-$3.53 & $-$3.26 & $-$3.05 & $-$5.34 & $-$4.85 & 99 &$-155.0$& ROE14b \\ 
SMSS\,J085924.06$-$120104.9 & $-$3.63 & $-$3.83 & $-$3.02 & $-$4.64 & $-$4.86 & 2387 &$-187.6$& JAC15 \\ 
      SDSS\,J1322+0123 & $-$3.64 & $-$3.15 & $-$3.39 & $-$4.88 & $-$4.94 & 2263 &105.0& PLA15a \\
    LAMOST\,J0126+0135 & $-$3.57 & $-$4.08 & $-$3.15 & $-$5.18 & $-$4.71 & 173 &$-256.0$& LI15a \\ 
SMSS\,J184825.29$-$305929.7 & $-$3.65 & $-$3.40 & $-$3.15 & $-$5.17 & $-$5.23 & 2471 &110.2& JAC15 \\ 
        HE~2318$-$1621 & $-$3.67 & $-$3.13 & $-$3.47 & $-$4.67 & $-$5.28 & 2080 &$-39.95$& PLA14a \\ 
           CS22952$-$015 & $-$3.68 & $-$4.52 & $-$3.57 & $-$4.52 & $-$5.37 & 968 &$-7.0$& ROE14b \\ 
       HE~0302$-$3417a & $-$3.70 & $-$3.22 & $-$3.15 & $-$5.05 & $-$5.80 & 2026 &121.7& HOL11 \\ 
        HE~2331$-$7155 & $-$3.70 & $-$2.34 & $-$2.48 & $-$4.53 & $-$4.58 & 2082 &210.6& HAN15 \\ 
        HE~0056$-$3022 & $-$3.72 & $-$3.52 & $-$3.39 & $-$4.68 & $-$5.23 & 1490 &29.7& ROE14b \\ 
        HE~0926$-$0546 & $-$3.73 & $<-$3.11 & $-$3.41 & $-$4.95 & $-$4.56 & 2637 &152.9& COH13 \\ 
        HE~1116$-$0634 & $-$3.73 & $-$3.65 & $-$2.91 & $-$5.99 & $-$5.54 & 2507 &116.09& HOL11 \\ 
        HE~0048$-$6408 & $-$3.75 & $-$4.03 & $-$3.35 & $-$4.57 & $-$5.23 & 2018 &$-33.99$& PLA14a \\ 
        HE~1012$-$1540 & $-$3.76 & $-$1.77 & $-$2.36 & $-$4.56 & $-$4.45 & 392 &225.6& ROE14b \\ 
           CS22960$-$048 & $-$3.78 & $-$3.44 & $-$3.19 & $-$5.91 & $-$5.50 & 815 &$-85.0$& ROE14b \\ 
SMSS\,J010651.91$-$524410.5 & $-$3.79 & $-$3.66 & $-$3.23 & $-$4.58 & $-$5.43 & 2355 &189.5& JAC15 \\ 
           CS30339$-$073 & $-$3.80 & $-$3.73 & $-$3.48 & $-$5.04 & $-$5.48 & 1451 &171.0& ROE14b \\ 
        BS16076$-$006    & $-$3.81 & $-$3.35 & $-$3.25 & $<-$5.30 & $<-$4.86 & 474 &203.12& BON09 \\ 
SMSS\,J004037.56$-$515025.2 & $-$3.83 & $-$3.92 & $-$3.24 & $-$4.82 & $-$4.84 & 2352 &72.9& JAC15 \\ 
           CS22189$-$009 & $-$3.85 & $-$3.62 & $-$3.48 & $-$4.81 & $-$5.44 & 222 &$-42.0$& ROE14b \\ 
           CS22963$-$004 & $-$3.85 & $-$3.58 & $-$3.51 & $-$5.04 & $-$4.70 & 236 &294.4& ROE14b \\ 
           CS22891$-$200 & $-$3.88 & $-$3.53 & $-$3.24 & $-$5.24 & $-$4.81 & 671 &137.7& ROE14b \\ 
        HE~1300+0157   & $-$3.88 & $-$2.54 & $-$3.50 & $<-$5.51 & $<-$4.74 & 2665 &73.4& FRE07a \\ 
SMSS\,J184226.25$-$272602.7 & $-$3.89 & $<-$4.18 & $-$3.27 & $-$5.66 & $-$5.16 & 2469 &$-247.0$& JAC15 \\ 
        HE~0945$-$1435 & $-$3.90 & $<-$1.87 & $-$3.76 & $<-$4.77 & $<-$3.87 & 2035 &121.8& HAN15 \\ 
        HE~1201$-$1512 & $-$3.92 & $-$2.72 & $-$3.66 & $<-$4.73 & $<-$3.81 & 2040 &237.2& YON13 \\ 
SMSS\,J005953.98$-$594329.9 & $-$3.94 & $-$2.73 & $-$3.32 & $-$5.04 & $-$4.58 & 2353 &375.3& JAC15 \\ 
      HE~1424$-$0241   & $-$3.96 & $<-$3.17& $-$3.57 & $<-$5.62 & $-$4.92 & 2646 &59.8& COH08 \\
           CS22172$-$002 & $-$4.00 & $-$3.77 & $-$3.68 & $-$5.07 & $-$5.08 & 257 &255.0& CAY04 \\ 
        HE~0057$-$5959 & $-$4.08 & $-$3.22 & $-$3.57 & $-$5.14 & $-$4.54 & 2019 &375.3& YON13 \\ 
        HE~2239$-$5019 & $-$4.20 & $<-$2.45 & $-$3.70 & $<-$4.75 & $<-$4.15 & 893 &368.7& HAN15 \\ 
           CS22885$-$096 & $-$4.21 & $-$3.81 & $-$3.57 & $-$6.16 & $-$6.05 & 701 &$-249.1$& ROE14b \\ 
 BD\,+44$^{\circ}$ 493 & $-$4.26 & $-$3.08 & $-$3.39 & $-$4.83 & $-$5.16 & 1192 &$-150.6$& ROE14b \\ 
SDSS\,J102915.14+172927.9 & $-$4.73 & $<-$3.68 & $-$4.70 & $<-$5.00 & ... & 2554 &$$-34.5$$& CAF11a \\ 
 CD\,$-$38$^{\circ}$ 245 & $-$4.50 & $<-$4.78 & $-$3.93 & $-$5.22 & $-$5.59 & 1509 &46.4& ROE14b \\ 
        HE~0557$-$4840 & $-$4.80 & $-$3.14 & $-$4.57 & $<-$5.77 & $<-$4.73 & 368 &211.94& NOR07 \\ 
        SD1313$-$0019  & $-$5.00 & $-$2.04 & $-$4.56 & $<-$5.28 & $<-$4.78 & 2262 &267.0& FRE15 \\ 
   HE~0107$-$5240      & $-$5.28 & $-$1.58 & $-$5.13 & $<-$5.80 & $<-$4.46 & 2021 &44.78& CHR04 \\ 
SMSS\,J031300.36-670839.3 & $<-$7.30 & $-$2.40 & $-$4.30 & $<-$6.70 & $<-$6.10 & 2136 &$-67.93$& KEL14 \\ 
        \hline
        HE~2319$-$5228 & $-$3.38 & $-$2.16 & $-$2.73 & $-$5.18 & $-$5.38 &  &292.8& this study \\ 
        HE~2340$-$6036 & $-$3.59 & $-$4.16 & $-$3.04 & $-$5.00 & $-$4.44 &  &212.4& this study \\ 
        HE~1310$-$0536 & $-$4.22 & $-$2.18 & $-$3.76 & $-$5.86 & $-$5.10 &  &188.3& this study \\ \hline
    \multicolumn{8}{l}{References are as follows: 
    LI15a: \citet{Li15},
    BAR05: \citet{Barklem05},
    ROE14b: \citet{Roederer14c},
    }\\
    
    \multicolumn{8}{l}{
    CAY04: \citet{Cayrel04},
    HAN15: \citet{Hansen15},
    LAI08: \citet{Lai08},
    PLA14a: \citet{placco14_bd},
   }\\
    
    \multicolumn{8}{l}{
    HOL11: \citet{hollek11},
     JAC15: \citet{Jacobson15},
COH13: \citet{cohen13},
NOR07: \citet{norris07},
}\\
    \multicolumn{8}{l}{
BON09: \citet{bonifacio09},
RYAN96: \citet{ryan96},
COL06: \citet{Col06},
YON13: \citet{Yong13a}, 
}\\
    \multicolumn{8}{l}{
KEL14: \citet{keller14},
FRE15: \citet{frebel15b}, 
CAF11a: \citet{caffau11},
COH08: \citet{cohen08}, 
}\\
    \multicolumn{8}{l}{
FRE07a: \citet{Frebel07}, MAR24: \citet{Mardini_CD-24}
}
    \end{tabular}
    \label{SASS}
\end{table*}

\section{Summary} \label{sec:conclusions}
In this paper, we studied a sample of metal-poor stars with low neutron-capture element abundances to gain insight into the origins of each of our program stars, and to investigate the potential connections of these halo stars to those in present-day UFDs and other dwarf galaxies. Our findings can be broadly summarized as follows: 

\begin{enumerate}

    \item High-resolution optical spectra of our target stars were obtained with Magellan/MIKE. Stellar parameters ($T_{\text{eff}}$, $\log g$, and $v_{\text{mic}}$) were determined as part of performing a detailed chemical abundance analysis. Depending on the element, we obtained abundances using either equivalent widths or spectrum synthesis. We obtained abundances of a total of 17 light and heavy elements. All our stars were found to have [Fe/H] between $-$2.9 and $-$4.3.

    \item The abundances of our six stars generally agree well with halo trends for most elements. However, for some elements (specifically Al, Ca, Ti, Co, Ni), our three low-Sr stars ([Sr/H] $< -4.5$; namely HE~ 1310$-$0536, HE~2340$-$6036, HE~2319$-$5228) show larger scatter and deviations from the main halo trend. In addition, two of our stars (HE~2340$-$6036 and HE~0104$-$5300) have C abundances that are far below average.

    \item The Sr and Ba abundances of our stars are extremely low, with our low-Sr program stars (square symbols in relevant figures) having values of [Sr/H] $<-$ 4.5 and [Ba/H] $<-$4.0. In terms of neutron-capture abundances, these three stars are highly similar to UFD stars (see Figures~\ref{fig:srba} and \ref{fig:ncap abundances}). The other three program stars (triangle symbols in relevant figures) are separated in a distinct cluster in chemical abundance space, following the general halo trend set by higher-metallicity stars. 
    
    Interestingly, this separation does not exist for Ba abundances, with all six program stars having [Ba/H] $\lesssim$ 4.0, and following the main trend. Therefore, Sr abundance is a telling sign of whether a star is similar to a UFD star, and by extension, old.

    \item The low-Sr program stars (square symbols in relevant figures) possess [Sr/Ba] ratios that do not reflect the main halo trend, but instead fall into the regime of UFD stars (see Figure~\ref{fig:srba}). We identify these three stars as having similar origins to the metal-poor stars observed in UFDs. We label these stars, and others with similarly low [Sr/H] values, as "SASS" stars. SASS stars are named as such because they likely originated in the early universe from Small Accreted Stellar Systems. Since these SASS stars are nearer to us compared to the very faint stars in surviving UFDs, identifying them in the halo allows for more detailed studies of accreted systems.

    \item From a kinematic analysis, we found that all program stars had extremely large and retrograde $V$ velocities. This retrograde motion is indicative of objects that were accreted by the Milky Way rather than formed within it. This supports our conclusion that the stars with the lowest Sr and Ba abundances are likely some of the earliest accreted objects.

    \item We searched the literature for additional SASS stars using the [Sr/H] $<-$4.5 cut and found 61 more such stars in the halo. A significant fraction of stars with [Fe/H] $<-$3 show these very low neutron-capture abundances. This fact supports the idea that the most metal-poor stars are also the oldest stars in our Galaxy (or at least the earliest stars that were accreted by the proto-Galaxy).

    \item The population of SASS stars is easily identifiable with low Fe and Sr abundances. This easy identification of ancient stars will ultimately allow insights into the earliest star-forming environments, the origins and histories of the surviving UFDs, as well as the earliest assembly phase of the Milky Way.

    \item The earliest systems were likely enriched by a very small number of supernovae whose yields were somewhat inhomogeneously mixed. This produced variations in light element abundances, including carbon abundances, from system to system. Neutron-capture elements were sparsely produced and/or diluted likely in different ways, with $r$-process nucleosynthesis playing a role. 

\end{enumerate}

Indeed, UFDs offer insights into the conditions of the early universe. In particular, they provide glimpses into early star formation environments, early chemical enrichment events, and even the earliest phases of the Milky Way's formation. However, their limited number of observationally accessible stars naturally limits their further exploration. Now, SASS stars may fill that gap and provide a new important line of access to the early cosmos. These stars, which now lurk in the Galactic halo, may shed further light on the nature and origins of the oldest stars in the universe.


\section*{Acknowledgements}
HDA, ASF, and CGF gratefully acknowledge support from the
MIT Undergraduate Research Opportunities Program Office. AF acknowledges support from  NSF grants AST-1716251 and AST-2307436. MKM and AF acknowledge partial support from NSF grant OISE 1927130 (International Research Network for Nuclear Astrophysics/IReNA). 

This work has made use of data from the European Space Agency (ESA) mission {\it Gaia} (\url{https://www.cosmos.esa.int/gaia}), processed by the {\it Gaia} Data Processing and Analysis Consortium (DPAC, \url{https://www.cosmos.esa.int/web/gaia/dpac/consortium}). Funding for the DPAC has been provided by national institutions, in particular the institutions participating in the {\it Gaia} Multilateral Agreement. This work made use of NASA's Astrophysics Data System Bibliographic Services. This research has made use of the SIMBAD database, operated at CDS, Strasbourg, France. This work made use of Astropy:\footnote{\url{http://www.astropy.org}} a community-developed core Python package and an ecosystem of tools and resources for astronomy \citep{astropy:2013, astropy:2018, astropy:2022}.

This work made use of general-purpose Python libraries, namely \textsc{numpy} \citep{Harris2020_numpy}, \textsc{matplotlib} \citep{Hunter:2007_matplotlib}, \textsc{pandas} \citep{pandas}, and \textsc{scipy} \citep{2020SciPy-NMeth}.

\section*{Data Availability}

The individual line measurements are provided as supplementary material. The reduced spectra can be obtained by reasonable request to the corresponding author.

\bibliographystyle{mnras}
\bibliography{sass} 

\bsp	
\label{lastpage}
\end{document}